\documentclass[11pt,sort&compress,fleqn]{cas-sc}

\usepackage{url}
\usepackage{xcolor}

\usepackage{verbatim} 	       
\usepackage{setspace}
\usepackage{mdframed}
\usepackage{enumitem}		   
\usepackage{subcaption}		   
\usepackage{hyperref}
\usepackage{placeins}		   
\usepackage{siunitx} 		   
\usepackage{numprint}		   
\usepackage{hhline}			   

\usepackage{natbib}
\def\tsc#1{\csdef{#1}{\textsc{\lowercase{#1}}\xspace}}
\tsc{WGM}
\tsc{QE}

\usepackage[demo]{adjustbox}
\usepackage{xcolor}
\usepackage{atbegshi}
\AtBeginDocument{\AtBeginShipoutNext{\AtBeginShipoutDiscard}}	
\PassOptionsToPackage{hyphens}{url}\usepackage{hyperref} 

\begin{document}
	
	\pagestyle{empty} 
	\begin{titlepage}
		\color[rgb]{.4,.4,1}
		\hspace{5mm}

		\bigskip
		
		\hspace{15mm}
		\begin{minipage}{10mm}
			\color[rgb]{.7,.7,1}
			\rule{1pt}{226mm}
		\end{minipage}
		\begin{minipage}{133mm}
			\vspace{10mm}        
			\color{black}
			\sffamily
			\LARGE\bfseries A computational framework  \\[-0.3\baselineskip] for nanotrusses:  \\[-0.3\baselineskip] input convex  \\[-0.3\baselineskip] neural networks approach     \\[-0.3\baselineskip] 
			
			\vspace{5mm}
			{\large {Preprint of the article published in \\[-0.4\baselineskip] European Journal of Mechanics - A/Solids (2023) }} 
			
			\vspace{10mm}        
			{\large Marko \v{C}ana\dj{}ija, Valentina Ko\v{s}merl, Martin Zlati\'{c}, Domagoj Vrtov\v{s}nik, Neven Munjas  } 
			
			\large
			
			\vspace{40mm}
			\vspace{5mm}
			
			\small
			\url{https://doi.org/10.1016/j.euromechsol.2023.105195}
			
			\textcircled{c} 2023. This manuscript version is made available under the CC-BY-NC-ND 4.0 license \url{http://creativecommons.org/licenses/by-nc-nd/4.0/}
			\hspace{30mm} 
			\color[rgb]{.4,.4,1} 
			\includegraphics[width=3cm]{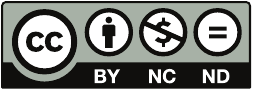}        
		\end{minipage}
	\end{titlepage}

\title [mode = title]{A computational framework for nanotrusses: input convex neural networks approach}  
\shorttitle{A computational framework for nanotrusses}    
\shortauthors{M. \v{C}ana\dj{}ija et al.}

\author[1]{Marko \v{C}ana\dj{}ija}[
       auid=000,
       orcid=0000-0001-6550-0258]
\cormark[1]
\ead{marko.canadija@riteh.uniri.hr}
\credit{Conceptualization, Methodology, Software, Validation, Visualization, Writing - original draft, Writing - review \& editing}

\affiliation[1]{organization={University of Rijeka, Faculty of Engineering, Department of Engineering Mechanics},
	addressline={Vukovarska 58}, 
	city={Rijeka},
	postcode={51000}, 
	country={Croatia}}
\affiliation[2]{organization={University of Illinois at Chicago, Department of Mechanical and Industrial Engineering},
		addressline={842 W. Taylor St.}, 
		city={Chicago},
		state={Illinois},
		postcode={60607-7022}, 
		country={USA}}	
\affiliation[3]{organization={Istrian University of Applied Sciences},
	addressline={Riva 6}, 
	city={Pula},
	postcode={52100}, 
	country={Croatia}}	
\author[1,2]{Valentina Ko\v{s}merl}
\credit{Software, Validation, Writing - review \& editing}
\author[1]{Martin Zlati\'{c}}
\credit{Software, Writing - review \& editing}
\author[1]{Domagoj Vrtov\v{s}nik}
\credit{Validation, Writing - review \& editing}
\author[3]{Neven Munjas}
\credit{Software, Writing - review \& editing}


\cortext[1]{Corresponding author}

\begin{abstract}
	The present research aims to provide a practical numerical tool for the mechanical analysis of nanoscale trusses with similar accuracy to molecular dynamics (MD). As a first step, MD simulations of uniaxial tensile and compression tests of all possible chiralities of single-walled carbon nanotubes up to 4 nm in diameter were performed using the AIREBO potential. The results represent a dataset consisting of stress/strain curves that were then used to develop a neural network that serves as a surrogate for a constitutive model for all nanotubes considered. The cornerstone of the new framework is a partially input convex integrable neural network. It turns out that convexity enables favorable convergence properties required for implementation in the classical nonlinear truss finite element available in Abaqus. This completes a molecular dynamics-machine learning-finite element framework suitable for the static analysis of large, nanoscale, truss-like structures. The performance is verified through a comprehensive set of examples that demonstrate ease of use, accuracy, and robustness.
\end{abstract}
 


\maketitle

\section{Introduction, motivation and outline}
The statement that the rapid development of nanotechnology enables all sorts of things became a clich\'{e} as an opening in publications on this subject. While it is true that nanotechnology represents a breakthrough in every branch of science and technology, the numerical apparatus that can handle the practical applications involving mechanics at the nanoscale is still not available. The problem is that the size of the structure becomes important, due to involvement of different physics than in conventional engineering. Nevertheless, standard macroscale solutions are still applied at the expense of lower accuracy. In order to adapt macroscale models to nanoscale problems, numerous theoretical models for truss- and beam-like nanostructures have been proposed. However, although the literature on this subject is particularly rich nowadays, their application to real-life problems demonstrating accuracy and conformity to molecular dynamics solutions or experiments is extremely rare. A detailed point-by-point discussion of the problems associated with such modeling of one-dimensional nanostructures such as carbon nanotubes (CNTs) can be found later in Sec.~\ref{sec_Rationale}. We now briefly note that these procedures critically depend on the existence of a material parameter describing the size effects, called the nonlocal parameter. Unfortunately, there are very few attempts to actually determine it, and this is not a simple thing to do.

In addition to the experimental determination of the nonlocal parameter, there are other possible approaches. Numerical calculations could be performed to obtain an estimate. The first approach is molecular structural mechanics (MSM) which is based on the beam finite elements representing the bonds between carbon atoms. The method is introduced in \cite{li2003structural}, applied to uniaxial tests of single-walled carbon nanotubes (SWCNTs), and demonstrated the size dependence of the Young's modulus with respect to diameter. Usually, the validity of the harmonic potential is assumed for interatomic bonds due to its simplicity, but there are other nonlinear approaches based on the Morse potential, see for instance \cite{meo2007molecular}. The ease of its implementation has triggered a series of papers on the subject, dealing with CNT or graphene as a two-dimensional structure. Notably, MSM analysis of the bending of SWCNT in \cite{Canadija17} showed that the nonlocal parameter is not a simple constant that can be used to fit size-dependent analytical models.

An alternative to MSM is molecular dynamics (MD). It is considered more realistic than MSM, but is more complex and computationally intensive. For carbon nanostructures, there are several potentials that the govern mechanical behavior, with the recently developed machine-learned potentials based on density functional theory considered the most accurate. In the case of carbon, the Gaussian approximation potential (GAP-20, \cite{rowe2020accurate}) seems to provide the accuracy of \textit{ab initio} simulations. However, practical applications with these potentials are much more computationally intensive than classical MD potentials and are therefore limited to smaller groups of atoms.

Thus, although the exact value of the nonlocal parameter is still an open issue, researchers are making significant efforts to obtain a tool for the analysis of more complex structures like nanotrusses and nanoframes. A logical choice is again the finite element method (FEM), typically a truss or beam element representing a one-dimensional nanostructure such as a carbon nanotube. Both static \citep{norouzzadeh2017finite, pinnola2022finite} and dynamic problems \citep{aria2019nonlocal,adhikari2021dynamic} are addressed. Unlike MSM, where each finite element represents a bond of a nanotube, here it represents an entire nanotube. These are then combined in a standard manner to form the required nanostructure. Since this type of finite element formulation relies on the one-dimensional nonlocal formulations described above and thus involves the nonlocal parameter, the same limited applicability issues apply to this approach as well.

Consequently, although analytical or FEM approaches seem to be a good idea, the question raised above about the exact value of the nonlocal parameter hinders their practical application. Motivated by this reason, the present research approaches the problem differently. In our previous research \citep{canadija2021deep,kosmerl2022}, we performed a comprehensive set of molecular dynamics simulations of uniaxial tensile tests of SWCNT at room temperature. It should be pointed out that an extensive analysis of the compressive behavior similar to \cite{canadija2021deep,kosmerl2022} is still lacking. Such an analysis, complementing the existing tensile analysis, could be used to obtain a dataset, and develop a deep learned neural network (NN) that can reproduce the uniaxial behavior. Unfortunately, straightforward application of a trained NN to nonlinear FEM is not possible. In short, since most practical cases of interest involve room temperature and MD simulations at room temperature involve stochastic thermal vibrations of atoms, these are also present in MD uniaxial tests. NNs capture these effects surprisingly well \citep{kosmerl2022}, but it turns out that this is actually a disadvantage for numerical implementations such as FEM. The tangential stress-strain modulus obtained in this way varies so much that convergence of the nonlinear FEM simulation becomes a problem. While it is possible to use physically-informed neural networks to smooth such oscillations \cite{haghighat2021deep}, such an approach requires that a specific constitutive model to which the data are fitted be incorporated into an NN. In the present case, such a constitutive model is not available, so an alternative approach should be taken.

Instead, it is still possible to include at least some of the physically observed behavior in the NN. For the problem at hand, this is the convexity of the potential used to define the constitutive model of SWCNT. The method is introduced in \cite{amos2017input}, where it is shown that a particular architecture of NN with some constraints on the choice of activation function and weights can indeed provide an output that is convex with respect to all or some input variables. Further, it will be shown later that such NNs can be integrable, which is an important extension of \cite{amos2017input}. Investigations in \cite{as2022mechanics,huang2022variational,bunning2021input}, used the possibility of integration to establish relationships between stress, strain and the potential. The NN is designed to enforce convexity of strain energy with respect to a part of the variables, while training is performed on the derivatives - stresses and strains. After training, integration can be performed to obtain the strain energy. A fully convex approach, where the strain energy depends solely on the deformation gradient, is presented in \cite{thakolkaran2022nn}. However, in contrast to the former references, in the latter reference the balance of the linear momentum is built into the loss function and thus requires displacements and support reactions for training instead of stress-strain pairs. Finally, all of the aforementioned research is related to feed-forward networks, but the same convexity methodology can be applied to recurrent neural networks as well \citep{chen2018optimal}.

To obtain a framework for the analysis of mechanical behavior of real-life nanotruss structures, the present study combines the advantages of the various methods described above. These include the accuracy of MD, the versatility of NN enriched with the convexity constraint, and the applicability of FEM. The above methods are combined to develop a truss finite element that can be applied to problems of this kind. The paper begins with a detailed discussion of the assumptions and limitations of the existing analytical and numerical models in Sec.~\ref{sec_Rationale}. This is followed by an extensive set of MD uniaxial tests, which are now extended to include the compressive behavior of SWCNTs in Sec.~\ref{sec_MD_model}, providing a dataset on which NN is trained. The convexity of NN with respect to strain is ensured by the specific choice of NN architecture, as mentioned earlier. Finally, the NN constitutive model is introduced into the finite element code Abaqus. This step was also not a straightforward one, and additional implementation problems concerning the correct value of the area of the cross-section had to be solved by introducing another NN. The extensive example section shows the possible applications of the newly developed MD-NN-FE framework, and some conclusions are drawn at the end of the paper.

\section{The Rationale for the MD-ML Truss Finite Element}
\label{sec_Rationale}
The standard approach in the literature dealing with mechanical modeling of one-dimensional nanostructures, in particular carbon nanotubes, is to postulate the existence of an analytical truss or beam model that can fully capture the mechanical behavior of a real structure. The central issue is the existence of forces that are nonlocal in nature (e.g. van der Waals forces). These forces are not relevant in the analysis of similar structures at the macroscale. For this reason, structures of this kind are said to be affected by small size effects. Such phenomena can be successfully addressed by the application of nonlocal mechanics. Nonlocal mechanics assumes that the state of stress (or strain) at a point is not only a function of the current strain (or stress) at the same point, but that the neighborhood also plays a role.

The present research focuses on the truss structures. For these, nonlocal mechanics usually relies on the landmark paper \citep{Eringen83}. To better motivate the rationale for the MD-ML based truss finite element presented here, a few basic points of Eringen are here recapitulated in the one-dimensional context. It is assumed that the stress field is a convolution:
\begin{equation} \label{eq_EringenStress}
\sigma= \int_L \phi \left( x - \xi \right)  E \varepsilon \left( \xi \right)  \mathrm{d} \xi, 
\end{equation}
where $\sigma, \varepsilon$ are stress and strain as usual, $E$ is the Young's modulus, and $\phi$ is an attenuation function or kernel describing the nonlocal influence of more closer and distant points. Typically, this is an exponential function that includes a scalar material parameter $\lambda$, which is known as the nonlocal parameter. The choice of kernel must respect certain conditions. In particular, as an extreme case where the structure is infinite, the classical macroscopic solution must be obtained, where the strain can be taken out of the integral, giving the normalization condition:
\begin{equation}
	\label{eq_EringenStress2}
	\begin{array}{c}
		\int_{\mathcal{B}_\infty} \phi (\left|\mathbf{x}-\boldsymbol{\xi} \right|,\lambda) \mathrm{d} V=1.
	\end{array}	
\end{equation} 
In other words, the kernel becomes the Dirac delta for a local case. However, the truss element under consideration is not infinite, so this condition cannot be strictly enforced. This leads to a number of paradoxes typically observed in beam models \citep{canadija2023book}. A better alternative, which is not prone to such shortcomings, is the stress-driven formulation introduced in \cite{Romano17}. This formulation introduces so-called constitutive boundary conditions to obtain a mathematically exact and differential formulation equivalent to the above integral formulation. In summary, the finite elements presented in the literature are based on formulations of this type.

Having presented the basics of analytical nonlocal formulations, we can point out some observations aimed at nanoscale structures:
\begin{itemize}
	\item What is the value of the nonlocal parameter $\lambda$? Any kind of application of these models critically relies on this question. However, there are only few attempts in the literature to determine $\lambda$. The lack of data can be explained as follows. Considering that nanoscopic structures are essentially composed of a discrete set of masses - atoms - one has to resort to MD. In MD, temperature manifests itself as stochastic thermal vibrations of atoms, with the associated kinetic energy directly affecting temperature. This also means that the structure is constantly mechanically excited by thermal vibrations, which triggers vibrations of the structure in essential modes. These vibrations are not small (see \cite{canadija2021deep} for an illustration) and make it almost impossible to distinguish between thermal vibrations and nonlocal effects (which are supposed not to be large).
	\item As can be clearly seen in the previous equations, when analyzing the stress/strain state in a point of a truss or beam, only the nonlocality with respect to the longitudinal coordinate is considered. However, strain at a point that does not have the same transverse coordinate also has an effect. Due to the high complexity of such issues, this is neglected in one-dimensional nonlocal frameworks.
	\item The vast majority of models assume material and geometric linearity.
	\item The more complex nanostructures are assemblies of simpler elements. What happens at the connection of two or more noncollinear nanotruss elements? Do neighboring elements also have a nonlocal effect on the element in question? This also raises the question of how the constitutive boundary conditions used in the stress-driven formulation can be implemented in such assemblies.
	\item It is also difficult to distinguish small size effects from the approximation regarding the ring cross-section of SWCNT.
\end{itemize}
Since the nonlocal one-dimensional finite elements are derived from such analytic models, they are inherently associated with these issues as well.

\section{Molecular Dynamics Simulations}
\label{sec_MD_model}
The present research starts from a dataset created during the development of a deep learning model to determine the mechanical properties of carbon nanotubes \citep{canadija2021deep} as obtained from uniaxial tensile MD tests of SWCNT. Subsequently, the same dataset was used to build another NN model capable of reproducing the tensile part of the stress-strain relationship of any single-walled carbon nanotube up to 4 nm in diameter \citep{kosmerl2022}. For the present purpose, the major properties of the model are briefly recapitulated, and the reader is referred to the references cited above for further details and discussion. At this point, it is important to emphasize that the tensile and compressive behaviors of SWCNTs differ, as observed in \cite{Chowdhury2012}. In particular, tension results in a nonlinear stress-strain curve, while compression is found to have an almost linear behavior that ends with the buckling of SWCNT. For this reason, the study in \cite{canadija2021deep} is now extended to include a similar series of compression tests with the same SWCNT configurations.

MD simulations were performed in LAMMPS \citep{Plimpton1995}. The modified AIREBO potential is chosen as the cornerstone of the model \citep{Shenderova2000}. Among the classical potentials, this one seems to be the most accurate \citep{qian2021comprehensive}. Although newer potentials derived by machine learning and relying on density functional theory have higher accuracy, they are also significantly slower compared to classical models, which hinders their application to the present case \citep{canadija2021deep}.

SWCNTs thus encompass all possible $(n,m)$ chiralities that provide diameters up to 4 nm, leading to 818 different configurations in ranging from (3, 3) to (51, 0), $n \ge m$. The length-to-diameter ratio was approximately 5:1. One end of a nanotube was fixed, while at the other end velocity was prescribed. All simulations were performed at a strain rate of 0.001 ps$^{-1}$. In preliminary investigations, other aspect ratios and strain rates were also considered, but it was found that these effects are so small at room temperature that the thermal vibrations mask their influence and cannot be reliably detected.

The calculation procedure involved the energy minimization of each configuration in the first step. Then, the temperature was increased from 0 to 300 K, followed by a uniaxial tension or compression test. The test takes place in a vacuum, i.e. without any kind of surrounding media. To reduce the randomness of the results due to thermal vibrations, the tension and compression of each configuration were simulated three times, resulting in a total of 4908 MD simulations. During the test, elongation, virial stresses, and diameter were monitored. The calculation of true stresses from virial stresses requires the specification of a geometry approximating the nanotube. For this purpose, a hollow cylinder was used whose cross-section was defined as a thin ring with a thickness of 0.34 nm. The diameter of the ring is obtained as in \cite{canadija2021deep}, but now smoothed with the NN as described in \ref{sec:app3}.

\begin{figure}
	\centering
	\includegraphics[scale=1.4]{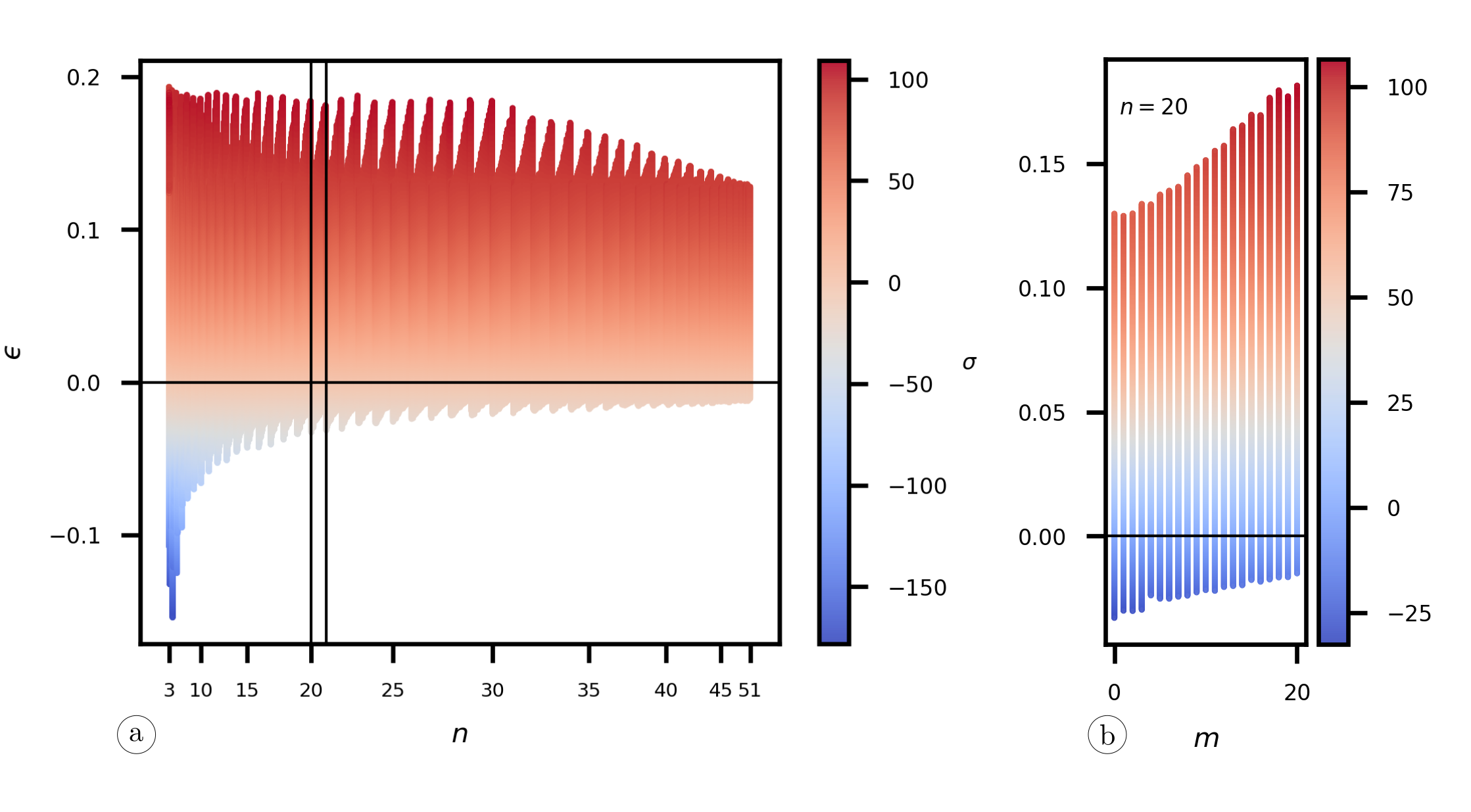}
	\caption{(a) The complete set of $\epsilon$ vs. $n$ results as obtained from MD. The color indicates the true stress level (GPa). Between two chirality parameters $n$ and $n+1$, the parameter $m$ increases. The vertical lines indicate the range of SWCNT $(20, 0)$ to $(20, 20)$ which is enlarged in (b).}
	\label{fig:stress2D}
\end{figure}

The true stress-true strain curve is obtained for each SWCNT. These curves are of particular interest to the current research and represent the dataset $\mathcal{D}$ consisting of $N(\mathcal{D})=\num{1469994}$ points. Comments and insights on these results are provided in Sec.~\ref{sec_ML_model} in comparison to machine learning results. At this point, only a general overview of the results is given in Fig.~\ref{fig:stress2D}. It is evident that the zigzag SWCNTs $(m=0)$ have the lowest extension and stress at fracture, while the armchair SWCNTs ($n=m$) have the highest extension and stress at fracture. The opposite is true for compressive behavior: zigzag SWCNTs can withstand the greatest shortening and stress.

Another issue can be noticed: up to $n=29$, it seems that strain at fracture, as well as tensile stress, are not affected by increasing $n$. A straightforward conclusion that for $n\ge30$ these two properties seem to become dependent on $n$ as well is not true. In particular, for the same $n$, a zigzag nanotube will have a smaller diameter than an armchair tube. For this reason, from $n=30$ armchair configurations have a diameter larger than 4 nm and were not included in the analysis, so that the largest armchair configuration is (29, 29). On the other hand, all zigzag SWCNTs up to (51, 0) were analyzed. The same progressively applies to chiral SWCNTs, so the last configurations analyzed were (50, 0), (50, 1), and (50, 2). Since the armchair SWCNTs have the largest strain at fracture and the zigzag ones the smallest, this is the source of the decrease in the maximum (and minimum) stresses/strains in Fig.~\ref{fig:stress2D}. Hence, the stress and strain at fracture depend on the chirality, while the diameter plays a minor role. The only exceptions are the smallest SWCNTs, and a discussion of this can be found in \cite{canadija2021deep,kosmerl2022}.

\begin{figure}
	\centering
	\includegraphics[scale=0.7]{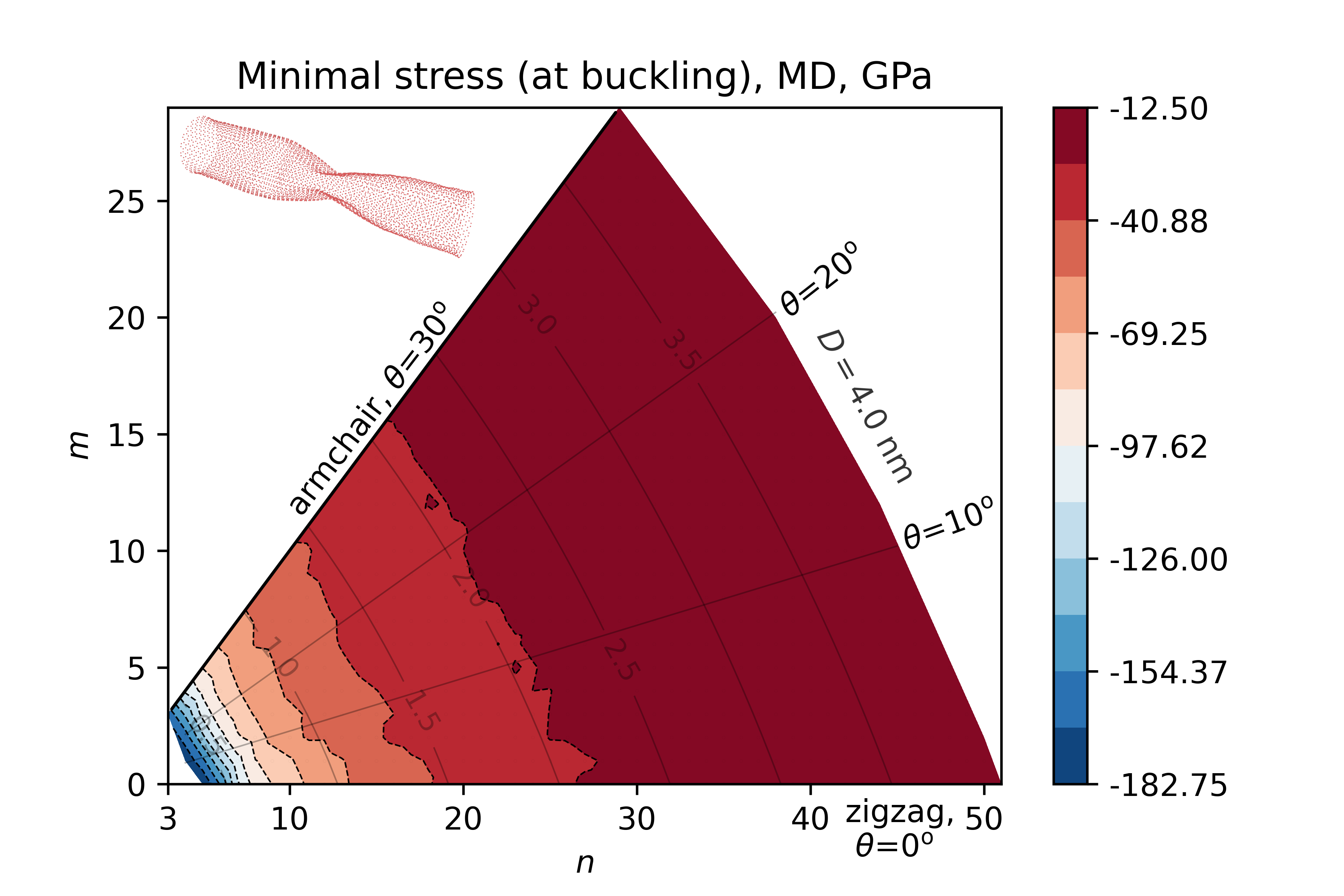}
	\caption{Minimum stress at buckling as obtained from MD simulations. The inset figure shows the failure mode of (20, 20) SWCNT (cf. Fig. 2 in \cite{buehler2004deformation}).}
	\label{fig:min_stress_MD}
\end{figure}

As for the compressive behavior, all SWCNTs failures can be attributed to buckling. The issue of CNT buckling is complex and beyond the scope of this article, so only the basic elements will be described and discussed. The minimum stresses at buckling are given in Fig.~\ref{fig:min_stress_MD}. A strong dependence on diameter can be seen, with SWCNTs with the largest diameter having the lowest stresses at buckling. This can be interpreted as follows. As pointed out in a review paper \citep{wang2010recent}, the compressive behavior of SWCNTs can be divided into shell-like, beam-like, and wire-like behavior, with the shell-like behavior observed at the smallest length-to-diameter ratio. The transition to the beam-like behavior is approximately at the aspect ratio $L/D=12.5$ \citep{buehler2004deformation}. Since in the present case the aspect ratio considered is $L/D=5$, all SWCNTs fail by shell buckling, see inset in Fig.~\ref{fig:min_stress_MD}.

This represents an important conclusion. As pointed out by \cite{yakobson1996nanomechanics} in shorter SWCNTs that buckle by shell buckling, the length of the SWCNT has virtually no effect on buckling. The critical strain is inversely proportional to the diameter, which also means that SWCNTs with the largest diameter have the lowest buckling strain/stress, as noted above. Thus, the results presented here for the critical buckling stress/strain are valid up to the aspect ratio $L/D=12.5$, i.e., until beam-like buckling begins to play a major role. This behavior is also observed for macroscopic cylinder-like structures \citep{nasa1975Astronautics}. For this reason, larger diameter SWCNTs are of limited use in mechanical problems involving compression. The problem could be mitigated by the surrounding media, but this is beyond the scope of the present research. Finally, the present results also confirm \citep{wang2010recent} that chirality correlates with buckling stress, such that the highest buckling stress is found in zigzag and the lowest in armchair SWCNTs, with chiral nanotubes following the same trend, Figs.~\ref{fig:stress2D}, \ref{fig:min_stress_MD}.

\section{Neural Network Constitutive Model of Single-Walled Carbon Nanotube}
\label{sec_ML_model}
The MD dataset $\mathcal{D}$ described in Sec.~\ref{sec_MD_model} contains nearly 1.5 million points in which the chirality indices $n$ and $m$ and the true strain are the input variables, while the output is the true stress. For implementation in the nonlinear truss finite element, a tangent modulus $\mathrm{d}\sigma/\mathrm{d}\epsilon$ is also required as an additional output. This dataset will serve as the basis for developing a neural network that can mimic the constitutive behavior of any axially loaded SWCNT with a diameter less than 4 nm. There is only one limitation to the model - SWCNT aspect ratio should be less than $L/D < 12.5$ to ensure that failure in compression is due to shell-like buckling, as discussed in Sec.~\ref{sec_MD_model}.

To this end, if one attempts to solve the issue with a straightforward application of classical machine learning (ML) networks  - either feed-forward or convolutional NN - one can obtain a very fine representation that even reproduces thermal vibrations in the stress-strain curves \citep{kosmerl2022}. Unfortunately, such an approach cannot be pursued if the finite element application is the goal. Namely, stochastic vibrations at the given simulation temperature lead to a loss of monotonicity in the stress-strain curve and, consequently, to a constant change of the tangent modulus. Convergence problems then arise in nonlinear finite element calculations. Another problem is that these models are prone to overfitting \citep{linka2023new}, while the underlying physical principles may be violated.

To solve these problems, a neural network can be informed about the fundamental physical constraints. For a start, the specific strain energy can be enforced to be a convex potential in the strain. Imposing the convexity property of strain energy ensures that the first derivative of strain energy with respect to strain - the stress-strain curve - is a non-decreasing monotonic function and the second derivative - the tangent modulus - remains positive.

In the research at hand, we do not require the strain energy to be convex in all input variables. Since the goal of the present research is to develop a neural network model for a very wide range of SWCNTs, the diameter and chirality angle should also be included in the formulation. However, the chirality indices also define the initial diameter and the chirality angle calculated by MD at a certain temperature (300 K). On the other hand, the use of the diameter of an unloaded SWCNT, as given by the well-known formula $D_\mathrm{th}=\frac{b\sqrt{3}}{\pi\sqrt{n^2+nm+m^2}}$, can only be considered as an approximation, since this equation is not related to temperature and cannot capture thermal changes in the diameter. In this sense, the diameter and chirality merely serve to identify a particular SWCNT configuration. Although it is not common to include parameters such as chirality as variables in the strain energy, this is the most convenient way to include all possible SWCNT configurations of interest and provide some sort of generalized potential. Thus, the diameter and chirality angle can alternatively be replaced by the chirality indices $n$ and $m$, and the convexity constraint on strain energy mentioned above should not be imposed on these variables. In summary, a neural network with a partially input convex neural network (PICNN) architecture is required, that is a neural network that is convex in strain energy with respect to the strain and nonconvex with respect to $n$ and $m$. A general concept for such networks is introduced in \cite{amos2017input} and serves as a starting point for this research. The main properties of the present approach are given in the rest of this section.

\subsection{Physical Constraints}
\label{sec_phy_constr}
The PICNN architecture can be used to provide the required convexity in the strain, where the output is the strain energy per current volume, generalized for all chiralities $\Psi(\epsilon, n, m)$. However, this energy is not part of the MD dataset, so it cannot be used directly and a different approach is taken. A similar technique is used in \cite{huang2022variational}, which is extended somewhat for the present purpose. In addition to dealing with convexity issues, the approach builds on the architecture of neural networks, which is integrable. For such networks, training is conveniently performed with partial derivatives $\sigma=\partial_\epsilon \Psi$ keeping $(\epsilon,n,m)$ as the input variables. Evaluation of the network with the trained parameters yields the strain energy and stresses as the first derivatives, while the second derivative corresponds to the tangent modulus. Having the above discussion in mind, this subclass of the PICNN is referred to as Partially Input Convex Integrable Neural Networks (PICINNs). Another advantage of such a PICINN formulation is thermodynamic consistency. That is, the validity of the second law of thermodynamics - for elastic processes, dissipation is zero, so stresses are defined as $\sigma=\partial_\epsilon \Psi$ - is inherently built into the NN. A similar approach has been used in \cite{as2022mechanics,huang2022variational,bunning2021input} for partial convexity, and elsewhere in the context of fully input convex neural networks, see for example \citep{thakolkaran2022nn, linka2023new} for models relating strain energy and deformation gradient invariants. In the end, training an NN that converges to a convex potential would be significantly less prone to overfitting \citep{linka2023new}.

Another physical constraint that should be included is the stress-free state at $\epsilon=0$. This assumes that the constitutive model ensures that rigid body motion does not cause stresses. We follow the same ideas used in \cite{as2022mechanics,huang2022variational}. The strain energy per unit of current volume modeled by the NN takes the following form:
\begin{equation}
	\label{eq_StrainEnergy}
	\begin{array}{c}
		\Psi^\mathrm{NN}(\epsilon, n, m)=\tilde{\Psi}(\epsilon, n, m)-\tilde{\Psi}(0, n, m)-\left. \partial_\epsilon \tilde{\Psi}(\epsilon, n, m)\right|_{\epsilon=0}  \epsilon.
	\end{array}	
\end{equation} 
Stresses are:
\begin{equation}
	\label{eq_Stress}
	\begin{array}{c}
		\sigma^\mathrm{NN}= \partial_\epsilon \Psi^\mathrm{NN}(\epsilon, n, m)=\partial_\epsilon \tilde{\Psi}(\epsilon, n, m)-\left. \partial_\epsilon \tilde{\Psi}(\epsilon, n, m)\right|_{\epsilon=0},
	\end{array}	
\end{equation} 
which must be used during training, since loss of accuracy for larger strains \citep{as2022mechanics} otherwise takes place. This ensures that the stress for the zero strain state vanishes, or in other words, the convex strain energy function has the minimum at $\epsilon=0$. It is also evident from Eq.~(\ref{eq_StrainEnergy}) that the addition of the term $\tilde{\Psi}(0, n, m)$ implies that $\Psi^\mathrm{NN}(0, n, m)=0$. Further, for $\epsilon\rightarrow \pm \infty$, Eq.~(\ref{eq_StrainEnergy}) gives $\Psi^\mathrm{NN}(\epsilon, n, m) \rightarrow \pm \infty$, i.e., to obtain infinite extension or compression, infinite energy is required. To summarize the above discussion, the specification of the strain energy in the form of Eq.~(\ref{eq_StrainEnergy}) introduces the physical constraints that the strain energy is non-negative, vanishes at $\epsilon=0$, requires $\sigma=0$ at $\epsilon=0$, and grows to infinity with $\epsilon\rightarrow \infty$.

The tangent operator is also required:
\begin{equation}
	\label{eq_tangent}
	\begin{array}{c}
		\partial_\epsilon \sigma^\mathrm{NN}= \partial^2_\epsilon \Psi^\mathrm{NN}(\epsilon, n, m)=\partial^2_\epsilon \tilde{\Psi}(\epsilon, n, m).
	\end{array}	
\end{equation} 

\subsection{Loss Function and Standardization}
\label{sec_loss}
With the above, a suitable loss function $\ell$ based on the mean squared error is:
\begin{equation}
	\label{eq_Loss}
	\begin{array}{c}
		\ell= \frac{1}{N(\mathcal{D}_\mathrm{d})} \sum\limits_{i=1}^{N(\mathcal{D}_\mathrm{d})} \left\lbrace \partial_\epsilon \tilde{\Psi}(\epsilon_i, n_i, m_i)-\left. \partial_\epsilon \tilde{\Psi}(\epsilon, n_i, m_i)\right|_{\epsilon=0} - \sigma_i \right\rbrace^2,
	\end{array}	
\end{equation} 
where $N(\mathcal{D}_\mathrm{d}) < N(\mathcal{D})$ is the size of the used part of the dataset (either training, validation or test dataset) $\mathcal{D}_\mathrm{d} \subset \mathcal{D}$, and $\lbrace\epsilon_i, n_i, m_i,\sigma_i\rbrace\in\mathcal{D}_\mathrm{d} $.

Further, it is known that standardization of the dataset variables helps with convergence in minimization. To this end, the dataset input $ \lbrace \epsilon_i, n_i, m_i \rbrace$ $\rightarrow \lbrace \overline{\epsilon}_i, \overline{n}_i, \overline{m}_i \rbrace $ was standardized:
\begin{equation}
	\label{eq_standard}
	\begin{array}{c}
		\epsilon=s_\epsilon \overline{\epsilon}+m_\epsilon\\
		n=s_n \overline{n}+m_n\\
		m=s_m \overline{m}+m_m,
	\end{array}	
\end{equation}  
where $s_{(\bullet)}$, $m_{(\bullet)}$ are the standard deviation and the mean of a particular input variable. Now, since the training is performed on the stresses Eq.~(\ref{eq_Stress}), then due to Eq.~(\ref{eq_standard})$_1$ and training with respect to the standardized $\overline{\epsilon} $ it must be:
\begin{equation}
	\label{eq_StandardSigma}
	\begin{array}{c}
		\sigma^\mathrm{NN}= \partial_\epsilon \Psi^\mathrm{NN}=\partial_{\overline{\epsilon}} \Psi^\mathrm{NN} \partial_\epsilon \overline{\epsilon} =
		\partial_{\overline{\epsilon}} \Psi^\mathrm{NN} \dfrac{1}{s_\epsilon}\\
		\text{or} \quad
		s_\epsilon\, \sigma^\mathrm{NN}= \partial_{\overline{\epsilon}} \Psi^\mathrm{NN}.
	\end{array}	
\end{equation} 
Since NN is now trained on the partial derivative of strain energy with respect to the standardized strains $\overline{\epsilon}$, the stresses that are part of the dataset are modified as $\overline{\sigma}=s_\epsilon \, \sigma$ and these values are used as labels during training. In the postprocessing step on the trained network, the inverse relation to Eq.~(\ref{eq_StandardSigma})$_2$ was used to obtain the stress predictions $\sigma^\mathrm{NN}$, while the tangent is calculated as:
\begin{equation}
	\label{eq_StandardSigma2}
	\begin{array}{c}
		\partial_\epsilon \sigma^\mathrm{NN}= \dfrac{1}{s_\epsilon^2}\partial^2_{\overline{\epsilon}\overline{\epsilon}} \Psi^\mathrm{NN}.
	\end{array}	
\end{equation} 
Standardization leads to a modification of the loss function Eq.~(\ref{eq_Loss}), which is eventually used in training:
\begin{equation}
	\label{eq_LossSt}
	\begin{array}{c}
		\overline{\ell}= \frac{1}{N(\mathcal{D}_\mathrm{d})} \sum\limits_{i=1}^{N(\mathcal{D}_\mathrm{d})} \left\lbrace \partial_{\overline{\epsilon}} \tilde{\Psi}({\overline{\epsilon}}_i, {\overline{n}}_i, {\overline{m}}_i)-\left. \partial_{\overline{\epsilon}} \tilde{\Psi}({\overline{\epsilon}},  {\overline{n}}_i, {\overline{m}}_i)\right|_{{\overline{\epsilon}}=-m_\epsilon/s_\epsilon} - {\overline{\sigma}}_i \right\rbrace^2.
	\end{array}	
\end{equation} 

Finally, the set of NN parameters $\lbrace \mathcal{W},\mathcal{B} \rbrace$, consisting of the set of all weight matrices $\mathcal{W}$ and the set of all bias vectors $\mathcal{B}$, is obtained as the solution of the optimization problem
\begin{equation}
	\label{eq_MinLoss}
	\begin{array}{c}
		\lbrace \mathcal{W},\mathcal{B} \rbrace= \arg \min\limits_{\mathcal{W}^*,\mathcal{B}^*} \overline{\ell}. 		
	\end{array}	
\end{equation} 
For minimization, the stochastic gradient descent algorithm Adam was used with a learning rate of $5\cdot 10^{-4}$.

\begin{figure}
	\centering
	\includegraphics[scale=0.7]{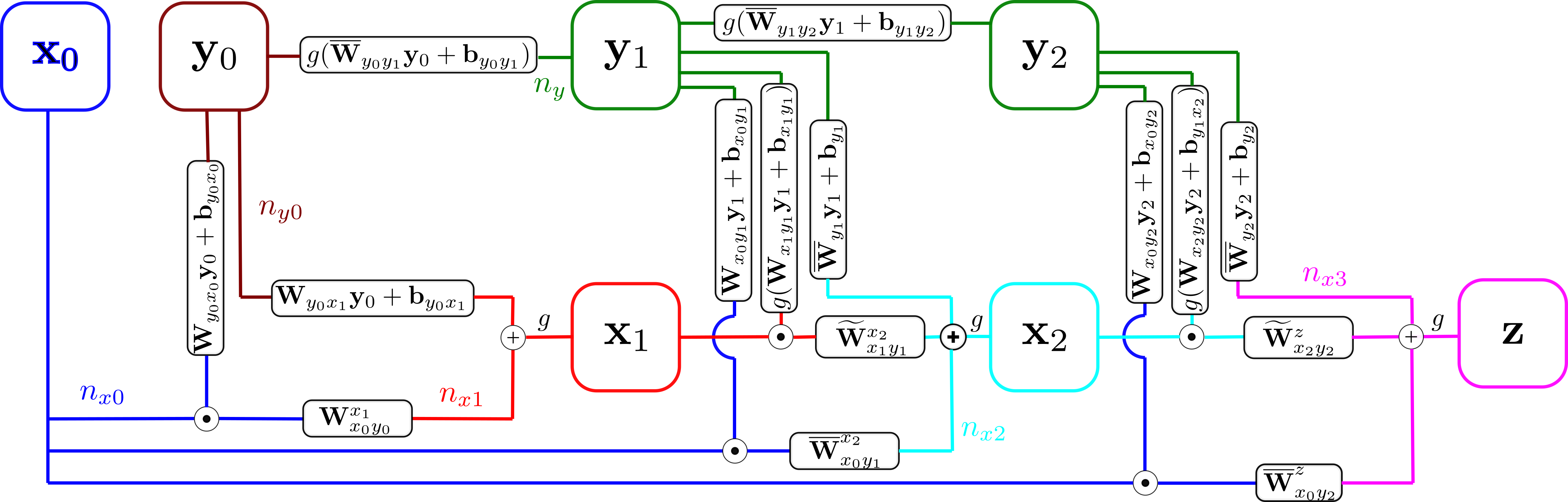}
	\caption{Neural network architecture referred to as pWN. Convex $\mathbf{x}_0=\left[ \epsilon \right] $ and non-convex $\mathbf{y}_0=\left[ n \quad m \right]^\mathrm{T}$ inputs, output $\mathbf{z}=\left[ \Psi^\mathrm{NN} \right]$. Colors indicate dimensionality of inputs/outputs, $n_{x0}=n_{x3}=1$, $n_{y0}=2$, $n_{y}=15$, $n_{x1}=n_{x2}=10$. $\overline{\mathbf{W}}$ indicate normalized weights in pWN neural network. $\widetilde{\mathbf{W}}$ denote normalized weights with non-negativity constraints.}
	\label{fig:NN}
\end{figure}

\subsection{Neural Network Architecture}
\label{sec_NN_arch}
The selected network architecture that conforms to these ideas is laid out in \cite{amos2017input, huang2022variational} and illustrated in Fig.~\ref{fig:NN}. The neural network consists of a combination of different dense layers and was developed using Keras and TensorFlow. In short, as pointed out in \cite{amos2017input}, imposing convexity of the output on a portion of the input variables relies on two aspects. For the present purpose, the reader should be reminded that the resulting function is convex if it is obtained as a (i) sum of non-negative convex functions and (ii) is a composition of non-decreasing convex functions.

In the context of NN, the first condition is fulfilled by setting non-negativity constraints on some of the weights, as discussed below. The second condition can be satisfied by an appropriate choice of activation functions. The resulting network is then of the PICNN type. However, in the present application, the network should also be integrable, i.e., belong to the PICINN subtype. This extension restricts the selection to such activation functions with derivatives that are common activation functions as well \citep{huang2022variational}. Among the standard activation functions, this leaves the softplus function $g(x)=\ln(1+\exp(x))$ with the logistic function as the first derivative $g'(x)=\exp(x)/(1+\exp(x))=1/(1+\exp(-x))$ as the only choice, while for the other, custom-made activation functions see \citep{linka2023new}. In this research, the softplus function is used in all cases.

The network is enriched by connecting the input layer directly to the hidden layers with the passthrough layers. This helps to reduce problems such as vanishing gradients, overfitting, and accuracy saturation \citep{he2016deep, thakolkaran2022nn} while maintaining the convexity of the network.

A formal proof of the above statements, somewhat extended from the proof originally given in \cite{huang2022variational}, is provided in \ref{sec:app1}.

\subsection{Weight Normalization}
\label{sec_Weight}
The present research extends the networks used in other works (especially in \cite{huang2022variational}) by means of the weight normalization, as proposed in \cite{salimans2016weight}, providing an increased convergence rate. In addition, the method also increases robustness, which was helpful in debugging of the NN in the development phase. The weight normalization relies on reparametrization of each weight vector $\mathbf{w}_i$ in the weight matrix $\mathbf{W}$ as $\mathbf{w}_i=\alpha_i \mathbf{n}_i$, where the scalar $\alpha_i$ and the vector $\mathbf{n}_i=\mathbf{w}_i/\left| \left| \mathbf{w}_i\right| \right|$ represent the new parameters. The scalar $\alpha_i$ then clearly represents the magnitude $\left| \left| \mathbf{w}_i\right| \right|=\alpha_i$, while the vector $\mathbf{n}_i$ describes the direction. Note that for the given non-negative weights, the weight normalization preserves the non-negativity constraint. Minimization is then performed for the new parameters $\alpha_i$ and $\mathbf{n}_i$ instead of for the weights $\mathbf{w}_i$.

Several possible choices for the weights that are normalized have been tested and are described below. The NN used in the finite element calculations was the pWN configuration, where only some of the weights were normalized in this way. These are denoted as $\overline{\mathbf{W}}$ in Fig.~\ref{fig:NN}. Although such normalization results in a somewhat longer training time, the main advantage is a better fit to the MD data manifested in a lower value of the loss function obtained after a certain number of epochs. Additionally, weights that were normalized in such a manner and simultaneously subjected to the non-negativity constraint (ii) to enforce convexity are denoted $\widetilde{\mathbf{W}}$. For more information about the non-negativity constraint, see \ref{sec:app2}.

\subsection{Training and Results}
\label{sec_train}
Configurations were tested in which the weights to be normalized were chosen differently. The configuration without weight normalization was used as a reference (referred to as noWN). The configuration 2WN used weight normalization only for the weights with the non-negativity constraints ($\widetilde{\mathbf{W}}_{x_1y_1}^{x_2}$, $\widetilde{\mathbf{W}}_{x_2y_2}^{z}$ in Fig.~\ref{fig:NN}). In the partial application of weight normalization (pWN), only some weights were selected (this is the case shown in Fig.~\ref{fig:NN}). In the case of full normalization (fWN), all weights were normalized. In all cases, the same training-validation-test split was used (ratio 60:20:20), but with different random initializations of the network parameters. The NN was trained for 500 epochs with a batch size of 1024. The Glorot uniform initializer was used.

The performance of the networks is presented in Tab.~\ref{tab:NNs}. Taking noNW network as the basis for comparison, it is clear that this configuration has the worst convergence, while the fNN network has the best. The same average validation loss of noWN network obtained after 500 epochs is achieved by the fNN network in only 138 epochs. On the other hand, the fNN configuration is the most expensive in terms of computation time. Thus, pWN seems to be the best choice that provides a tradeoff between convergence quality and training time. It also provided the lowest test loss (pWN ID 2) and is therefore selected for the finite element implementation.

\begin{table}
	\begin{center}
		{		\footnotesize
			\begin{tabular}{c|c|c|c|c|c|c|c}
				\hhline{=|=|=|=|=|=|=|=}
				& ID 1 & ID 2 & ID 3 & ID 4 & ID 5 & avg & $t/t_\text{noWN}$ \\
				\hline
				noWN (train.)& 4.478	& 7.299$^{a}$ & 4.514 & 2.873 & 2.360 & 4.305   & \\
				noWN (val.)  & 4.860	& 6.650$^{a}$ & 4.412 & 3.032 & 2.433 & \underline{4.277}$^{b}$  &1.00 \\
				noWN (test)  & 4.936	& 6.766$^{a}$ & 4.481 & 3.056 & 2.447 & 4.337 & \\
				\hline
				2WN (train)&  1.337 & 0.820 & 1.424 & 1.991 & 1.725 & 1.459  &\\
				2WN (val.) &  1.241 & 1.038 & 1.363 & 1.954 & 1.745 & 1.468  & 1.16\\
				2WN (test) &  1.250 & 1.048 & 1.362 & 1.948 & 1.729 & 1.467  & \\		
				AE 	& 329 & 184 & 89 & 282 & 228 & 222.4  &\\
				\hline
				pWN (train)& 1.523 & \textbf{0.739}$^{c}$ & 0.955$^{d}$ & 1.775 & 2.283 & 1.455  & \\
				pWN (val.) & 1.566 & \textbf{0.784}$^{c}$ & 0.996$^{d}$ & 1.688 & 2.274 & 1.462  &1.50\\
				pWN (test) & 1.566 & \textbf{0.795}$^{c}$ & 1.009$^{d}$ & 1.690 & 2.271 & 1.466  &\\	
				AE &132 & 91 & 155 & 187 & 382 & 189.4 & \\
				\hline
				fWN (train)& 0.880 & 1.137 & 0.673 & 1.250 & 0.926 & 0.973 &\\
				fWN (val.) &	0.818 & 1.108 & 0.816 & 1.182 & 0.943 & 0.973 &2.24 \\
				fWN (test) & 0.831 & 1.115 & 0.819 & 1.199 & 0.953 & 0.983 & \\
				AE & 161 & 74 & 150 & 150 & 157 & 138.4 &\\
				\hhline{=|=|=|=|=|=|=|=}
			\end{tabular}
		}	
		\caption{Training, validation, and test losses $\cdot10^{-4}$ after 500 epochs for 5 trainings. Abbreviations: noWN/2WN/pWN/fWN - without/two weights/partial/full weight normalization on dense layers, AE - average number of epochs until average validation loss on noWN network $4.277\cdot10^{-4}$ is reached ($^{b}$), $t$ time, $t_\text{noWN}$ average training time for noWN network. $^{a}$ minimum loss is reached at epoch 398, $^{c}$ best network, $^{d}$ minimum loss is reached at epoch 284. In all other cases, best loss is achieved at epoch 500.}
		\label{tab:NNs}
	\end{center}
\end{table}

\begin{figure}
	\centering
	\includegraphics[scale=1.0]{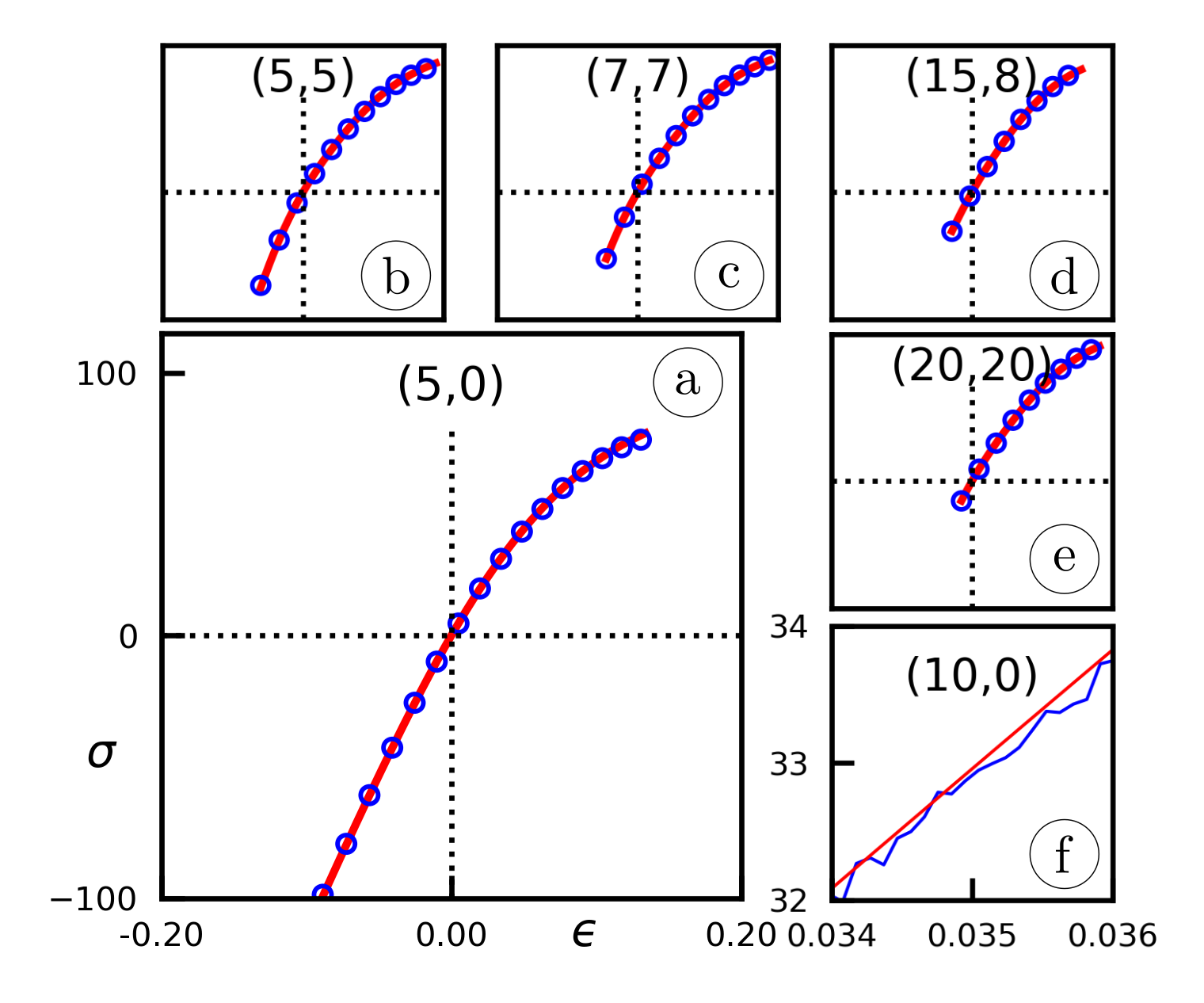}
	\caption{(a)-(e) comparison of $\sigma$ vs. $\epsilon$ curves for selected CNTs as obtained from the NN (red) and MD (blue); $\epsilon \in \left[ -0.2, 0.2 \right]$,  $\sigma \in \left[ -100, 115 \right] $ GPa for all axes. (f) A detail of $\sigma$ vs. $\epsilon$ curve for (10, 0) CNT.}
	\label{fig:se}
\end{figure}

A comparison of the stress-strain curves as obtained from ML with the curves from MD is shown in Fig.~\ref{fig:se}. Visual differences can be observed only with very close inspection. Examination of Fig.~\ref{fig:se}f also demonstrates the effect of the convexity constraint. While in the MD data stochastic fluctuations and loss of convexity are clearly visible, the NN curve is a monotonic curve.

\begin{figure}
	\centering
	\includegraphics[scale=0.6]{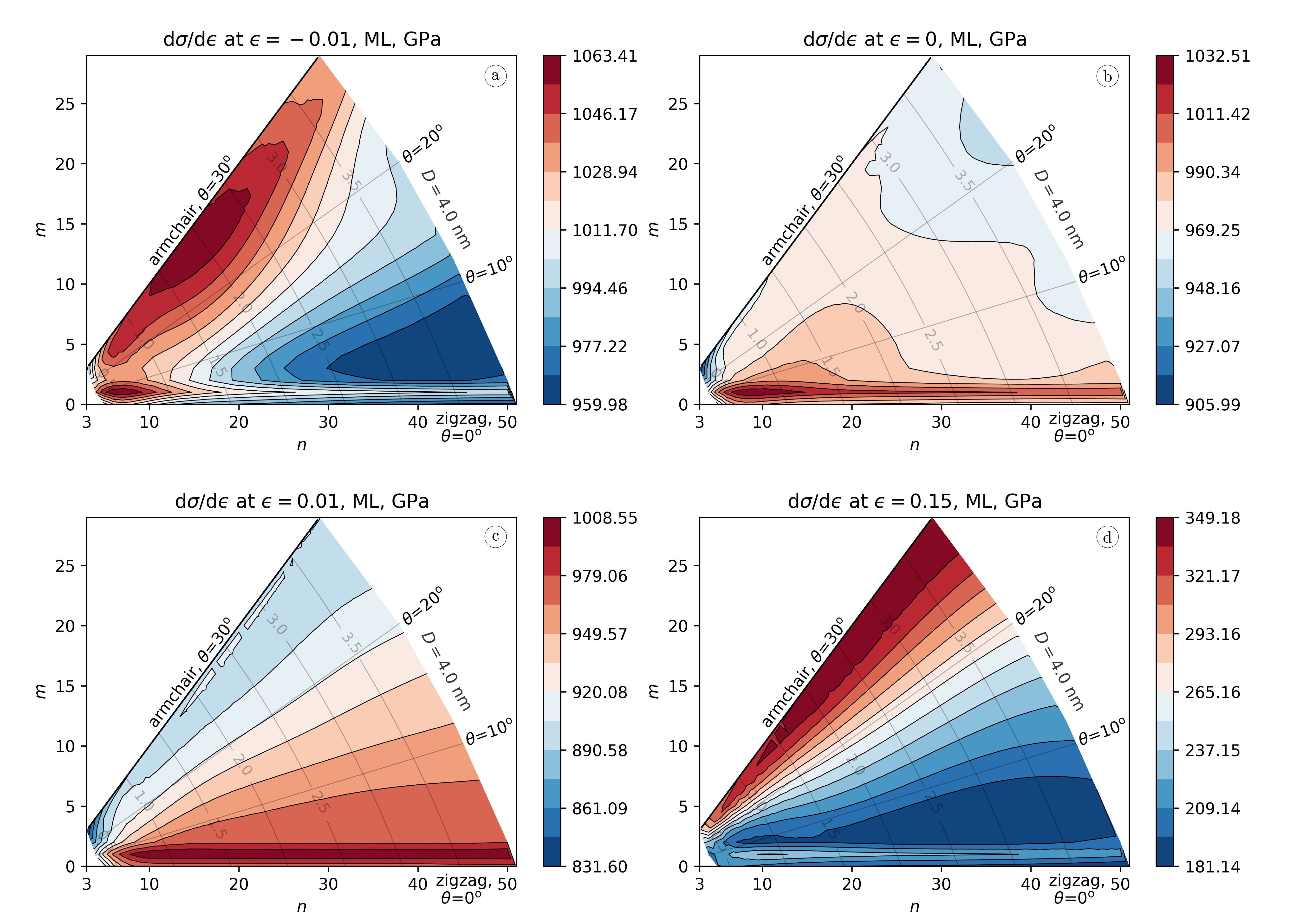}
	\caption{Tangent modulus $\mathrm{d}\sigma/\mathrm{d}\epsilon$ at $\epsilon=\lbrace-0.01, 0, 0.01, 0.15\rbrace$ as obtained by the NN.}
	\label{fig:YMeps0}
\end{figure}

The tangent modulus $\mathrm{d}\sigma/\mathrm{d}\epsilon$ is shown in Fig.~\ref{fig:YMeps0} for several different strain levels $\epsilon=\lbrace-0.01, 0, 0.01, 0.15\rbrace$. Of particular interest is the modulus for the strain-free configurations. The distribution of the latter property is already presented in \cite{canadija2021deep}, which at first glance looks somewhat different from the present results. The difference can be explained by considering that the results in \cite{canadija2021deep} were obtained from tensile-only simulations for the same set of SWCNTs as here, in contrast to the present results, which were obtained from compression and tension tests.

Comparing the tangent modulus obtained for slight compressive strain in Fig.~\ref{fig:YMeps0}a with the same absolute level of tensile strain in Fig.~\ref{fig:YMeps0}c, reveals the opposite trends. While for tensile loads the largest moduli are obtained for configurations close to zigzag SWCNTs and the lowest for armchair, exactly opposite is true for compressive loads (with exceptions for smaller diameter chiral CNTs close to zigzag configurations). This in turn has an averaging effect on the moduli in the strain-free state, which is visible in Fig.~\ref{fig:YMeps0}b and explains the differences from \citep{canadija2021deep}. Otherwise, the dominant influence of chirality is observed at all strains.

Interestingly, the largest moduli are obtained at smaller strains for small-diameter SWCNT configurations near zigzag SWCNTs, while these moduli drop most sharply as deformation proceeds and even assume the smallest values at high strains, Fig.~\ref{fig:YMeps0}d. As for the size-dependent behavior, i.e., the dependence on diameter observed for smaller diameters at all strains, it should be kept in mind that the calculations involve approximation of the irregular polygon with carbon atoms in the vertices to a circle to obtain the cross-section used to determine the stresses. Therefore, it is difficult to confirm unambiguously that the size effects in the SWCNT moduli exist or are caused by the cross-section approximation or both. On the other hand, the influence of chirality is clearly visible.

\begin{figure}
	\centering
	\includegraphics[scale=1.0]{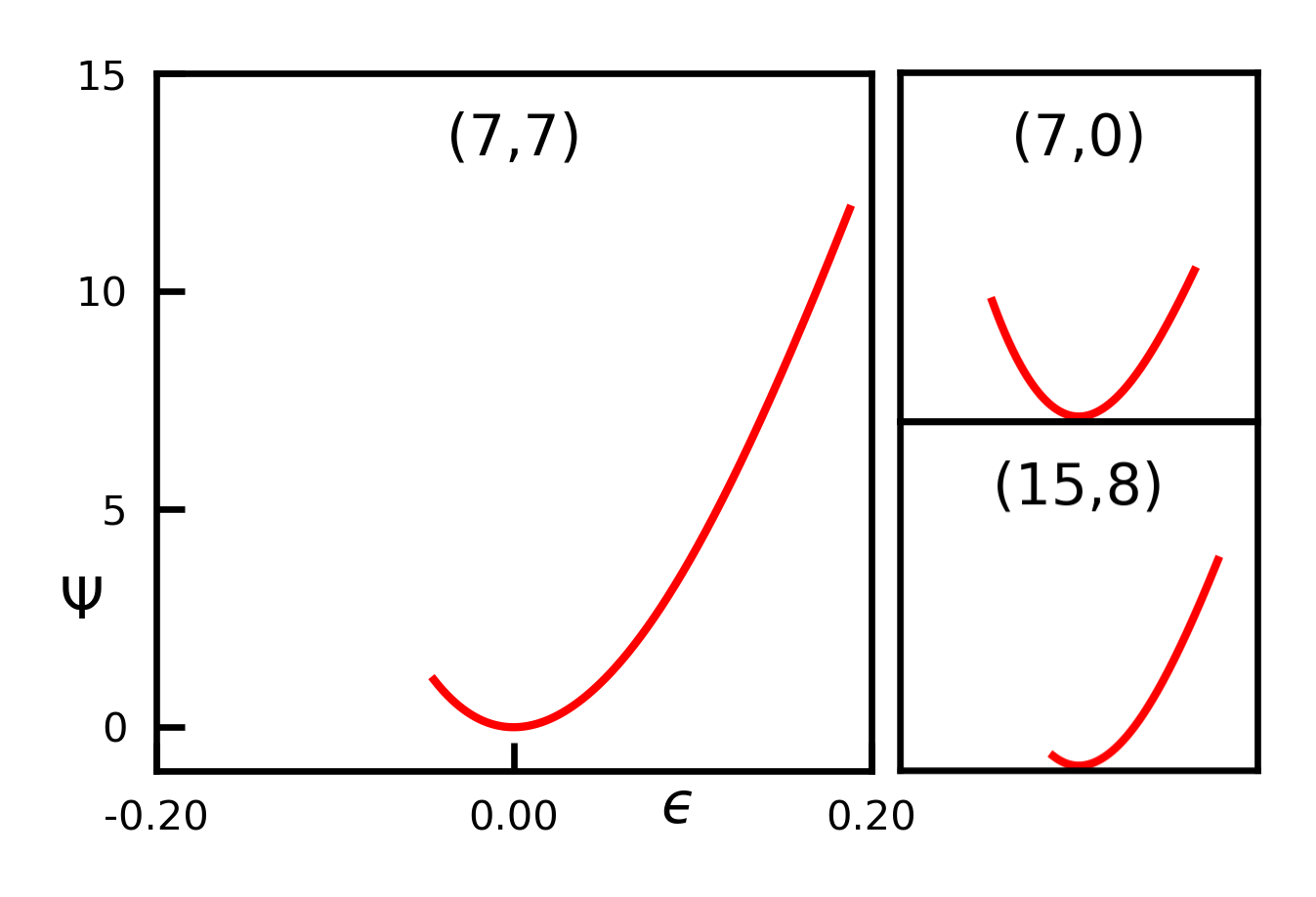}
	\caption{Specific strain energy (GPa) vs. true strain for selected SWCNTs, as predicted by the NN.}
	\label{fig:psi}
\end{figure}

Fig.~\ref{fig:psi} gives typical changes in specific strain energy vs. true strain. Again, differences in constitutive behavior in tension and compression can be seen. In particular, the energy is not symmetric about the $\epsilon=0$ axis, and a portion of the energy related to the compressive domain is shortened as the SWCNT diameter increases.

\begin{figure}
	\centering
	\includegraphics[scale=1.0]{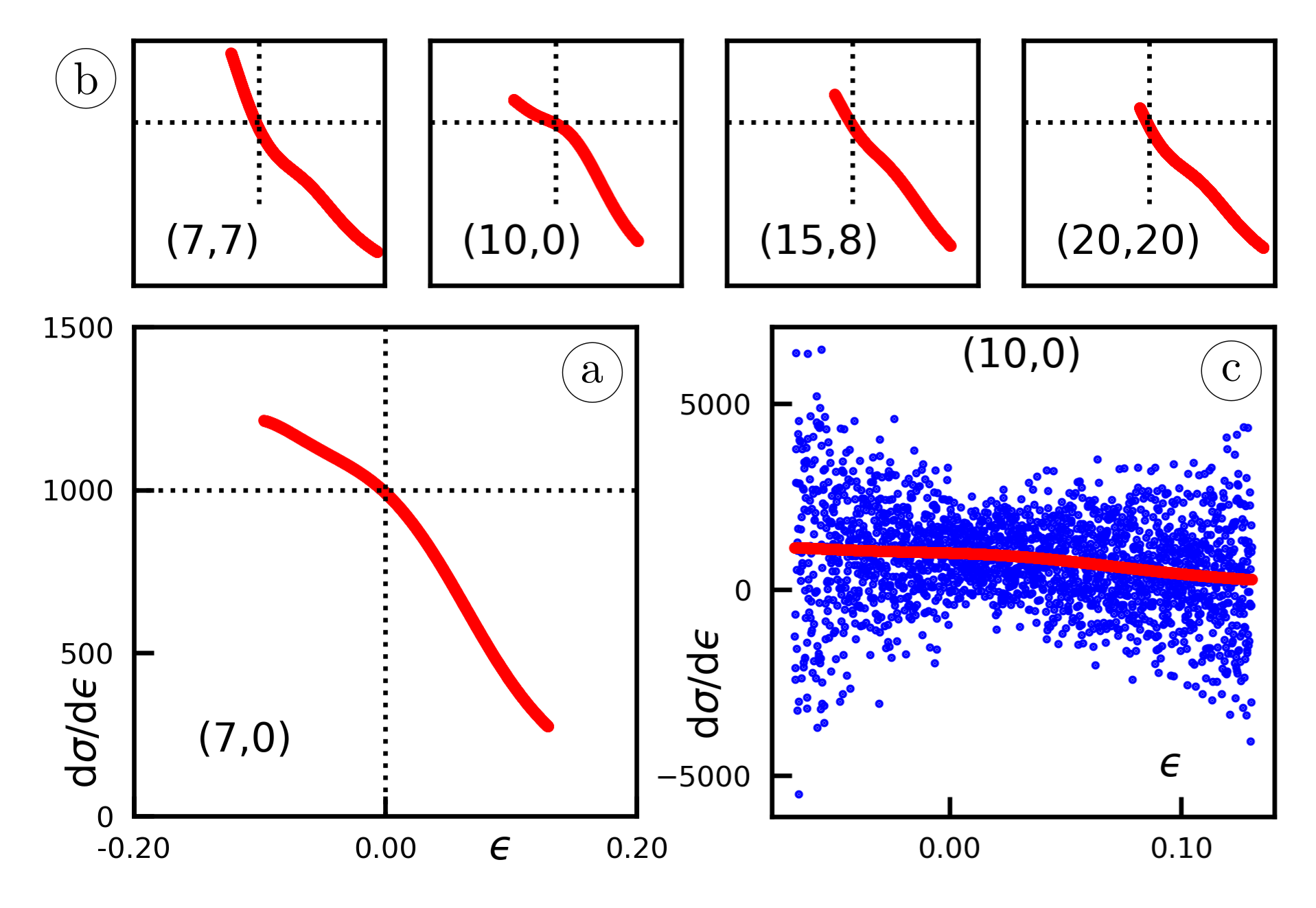}
	\caption{Typical $\mathrm{d}\sigma/\mathrm{d}\epsilon$ vs. $\epsilon$ curves (a), (b) for selected CNTs as obtained from the NN; in all figures $\epsilon \in \left[ -0.2, 0.2 \right]$,  $\mathrm{d}\sigma/\mathrm{d}\epsilon \in \left[ 0, 1500 \right] $ GPa. Dotted lines drawn at $\epsilon=0$ and $\mathrm{d}\sigma/\mathrm{d}\epsilon=1000$ GPa; same limits used on all axes; (c) comparison of $\mathrm{d}\sigma/\mathrm{d}\epsilon$ vs. $\epsilon$ obtained by the NN (red) and first order approximation $\Delta\sigma/\Delta\epsilon$ from MD (blue).}
	\label{fig:dsde}
\end{figure}

\section{Finite Elements and Implementation}
\label{sec_abaqus}
In Sec.~\ref{sec_Rationale} several issues are raised regarding the classical nonlocal approach in one-dimensional nanostructures. Perhaps the most crucial point is the exact value of the nonlocal parameter. While it might be possible to determine it at 0 K in the absence of stochastic thermal oscillations, but for most practical applications involving structures assembled of individual atoms, such as carbon nanotubes at room temperature, this is rather difficult, if possible at all. This conclusion is supported by the fact that the nonlocal effects are small, so thermal vibrations may hinder meaningful extraction of data used to calculate the nonlocal parameter. It should be emphasized that in the present analysis isolated carbon nanotubes are considered, without any influence of the surrounding medium, either solid or fluid. When the structure is immersed in another medium, stochastic thermal vibrations are still present, but the self-excited vibrations that these are triggering will be damped. In theory, this should be a step toward easier determination of the nonlocal parameter, but other questions arise: how to model the interactions between the nanotube and the medium, what happens to the nonlocal parameter when a different medium is used, and so on.

In strong contrast to the above, skipping the need to determine the nonlocal parameter is advocated in the present research. This is done by coupling the results of MD simulations with the nonlinear truss finite element. Deep learning is used as an interface between these two numerical techniques, resulting in a kind of hybrid finite element. A constitutive material model is not required, so no assumptions about material behavior need to be made (apart from convexity of strain energy). The role of the constitutive model will be taken by a neural network. This is a significant advantage over the classical approach. 

The above discussion also implies that any kind of material nonlinearity and nonlocality is built into the formulation. Conversely, the one-dimensional nonlocal models available in the literature almost exclusively use the initial constant Young's modulus, whereas the present approach uses a tangent modulus appropriate to the current strain. Geometric nonlinearities are handled in the same way as in classical finite elements.

Nevertheless, there are two issues that are not fully addressed even in this approach. In particular, the behavior at the joint where several members are joined together involve a small rotational influence of the other member atoms on another member. This can be solved by introducing some sort of stiffness into the joints, which is known in structural engineering as semi-rigid joints \citep{elsheikh1993}, and requires the use of beams rather than trusses. In the present case, this would require a series of MD investigations performed on different types of joints. Arguably, as visible in the Ex.~\ref{ex_3DMD_FEM_Truss} the effect is not significant, but a more extensive investigation might provide more evidence. The last point is the problem of the shape of the cross-section mentioned above, which is also open to possible improvements. Most likely, the issue can be addressed by directly relating the force and dilatation of a member through another deep neural network. However, this could be much more difficult to implement than resorting to stress-strain relationships.

Although the underlying concept is not too complex, the Abaqus implementation is not exactly straightforward. Everything works well for small deformations, but larger deformations require further modification. In particular, geometric nonlinear truss elements are considered incompressible in Abaqus. While this works well in the case of metal plasticity or incompressible hyperelasticity, in the present case the area of the cross-section is increasingly miscalculated with increasing strain. This in turn affects the accuracy of the nodal internal forces. To address this issue, another quantity is introduced - the rescaled stress, which is calculated as $\tilde{\sigma}=\xi \sigma$, where $\xi=D L/(D_0 L_0)$, $D_0, L_0$ and $D, L$ are the initial and current diameter and length of the SWCNT, respectively. Likewise, the tangent operator is replaced by $\mathrm{d} \tilde{\sigma}/\mathrm{d} \epsilon =\xi^2 \mathrm{d} {\sigma}/\mathrm{d} \epsilon$. This requires knowledge of the functional dependence $D=f(n,m,\epsilon)$, for which another neural network is developed based on MD data and described in \ref{sec:app3}.

Considering the above discussion and the results described in Sec.~\ref{sec_MD_model} and Sec.~\ref{sec_ML_model}, this can be implemented relatively easily in most finite element codes that have the ability to define user subroutines for the constitutive behavior. As mentioned earlier, this particular implementation was done in Abaqus-Simulia using the UMAT subroutine. Within the UMAT subroutine, the input data was passed to the Python code containing the trained neural network, which returned the true stress, the corresponding tangent operator, the initial diameter, and the current diameter of the SWCNT. In summary, the approach builds on the existing two node truss finite element in the Abaqus code and combines it with the machine-learned constitutive model based on the MD simulations. As such, it can be considered a hybrid MD-ML-truss finite element. The implementation turned out to be quite accurate and robust, as it will be demonstrated in the following section.

\section{Examples}
\label{sec_examples}
In this section, the performance of the newly developed framework is demonstrated on five problems: uniaxial tension/compression test on simple rods, a tetrahedron loaded in three different ways, a three-dimensional space truss, and two examples involving a hypothetical metamaterial design: size effect analysis and the second-order octahedron of octahedra of an octahedra lattice. In all examples, the Abaqus T3D2 rod element was used.

\subsection{MD to FEM Comparison 1: Uniaxial Tensile and Compressive Test of SWCNTs}
\label{ex_2UniAx}
The problem of uniaxial loading of a simple rod is solved first. One end of the rod is held fixed while the load is applied to the other end. In the MD simulations the velocity is used as the load, while in FEM the axial force is prescribed. Both tension and compression were simulated as two separate loading cases. Three CNT configurations were tested and their geometric properties and loading are listed in Tab.~\ref{tab:Ex1} below: zigzag (7, 0), chiral (15, 8), and armchair (20, 20). The length $L$ at the end of equilibration, as obtained from MD simulations, was assigned to the truss SWCNT finite element. As for the diameter, the following procedures were used. As described in Sec. \ref{sec_MD_model}, the diameters were recorded during MD simulations and used directly in the calculation based on the virial stresses that are here referred to as MD results. As for the FEM formulation, constant fluctuations in diameter $D(\epsilon)$ could lead to convergence problems, so an additional NN was trained to provide current $D$ as described in \ref{sec:app3}. Evaluation $D(0)$ provided the initial values for the diameters. To solve the incompressibility issue mentioned in Sec.\ref{sec_abaqus}, the latter NN was used to obtain the correct value of the current diameter by rescaling the stresses and tangent operators.

\begin{figure}
	\centering
	\includegraphics[scale=1]{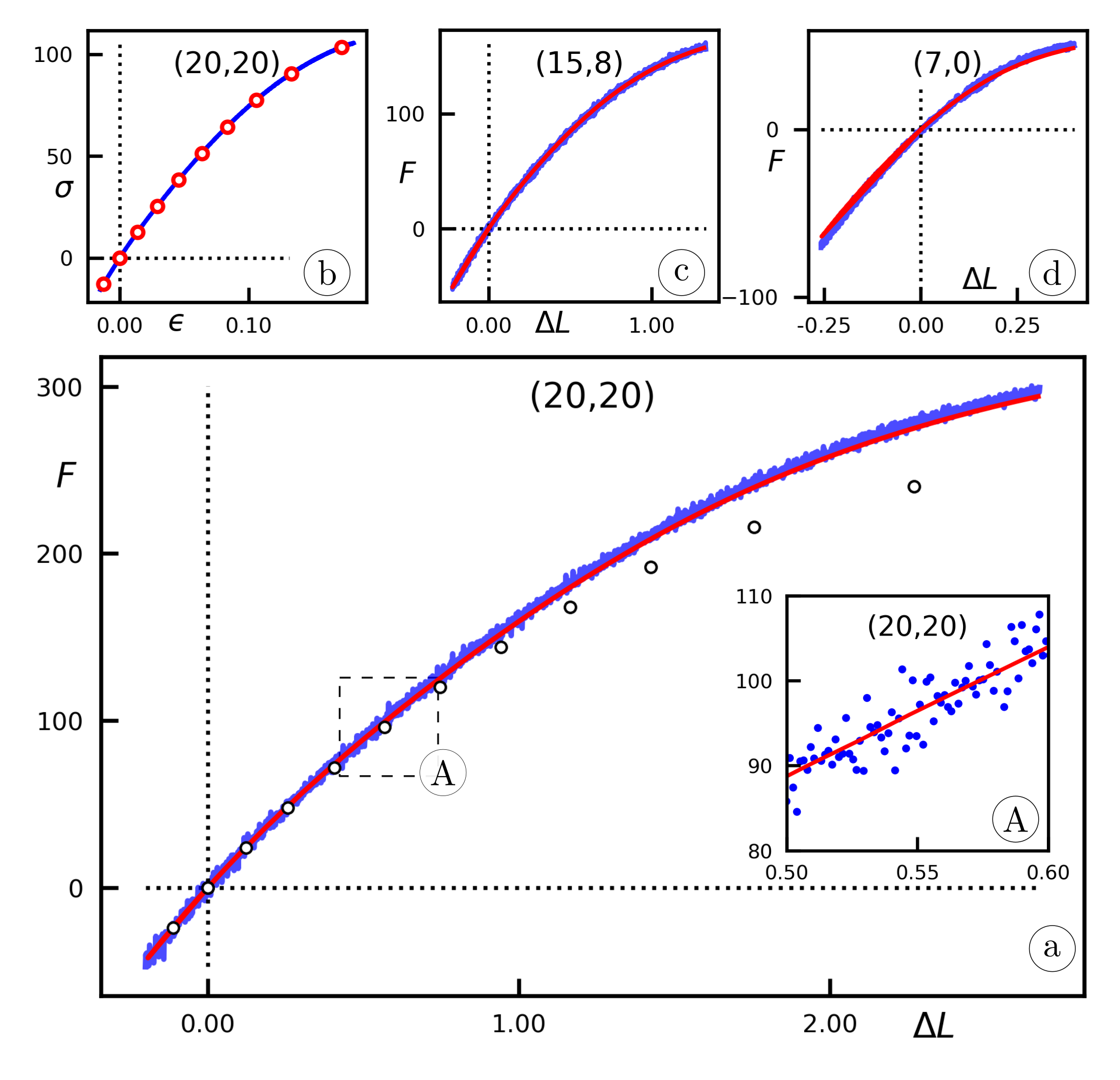}
	\caption{Uniaxial tests of SWCNTs - comparison of results from MD (blue) and FEM (red). (a) $F-\Delta L$ armchair (20, 20), (b) $\sigma$ - $\epsilon$ armchair (20, 20), (c) $F-\Delta L$ chiral (15, 8), (d) $F-\Delta L$ zigzag (7, 0). The black circles in (a) represent results obtained with diameters computed with an incompressible formulation available in Abaqus. Inset A shows the convexity of the NN formulation. Forces in nN, displacements in nm, stresses in GPa.} 
	\label{fig:uniaxial}
\end{figure}

The obtained results for the MD and FEM agree well for all three CNT configurations, see Fig.~\ref{fig:uniaxial}. SWCNTs (15, 8) and (20, 20) are typical representatives of the obtained curves. The zigzag SWCNT (7, 0) shows a slight error at the end of the compression phase. Similar results are also observed for a few other curves, which are attributed to a sharp increase in the maximum compressive stresses in SWCNTs with the smallest diameter. Results obtained by the incompressible formulation as provided by Abaqus, Fig.~\ref{fig:uniaxial}a, show that this can be a viable simplification for small strains. For larger strains, significant differences in MD and NN appear, supporting the argument for developing separate NN for the current diameter approximation. The inset in the same figure shows that the developed NN successfully enforces the convexity of the formulation. Finally, it should be noted that the NN does not perform well outside the range of input variables for which it was trained, as is common in these problems.

\begin{table}
	\caption{Geometric properties of uniaxially loaded CNTs in Ex. \ref{ex_2UniAx}}
	\centering
	{		\footnotesize
		\begin{tabular}{c|r|c|c|c}
			\hhline{=|=|=|=|=}
			CNT 			 &  Length &  Diameter & Cross-section & Axial \\
			configuration 	 &   (nm) &  (nm)  & area (nm$^2$) & force (nN)\\	
			\hline
			zigzag (7, 0)     &   2.8507 & 0.556  & 0.5940 & -75/60 \\ 
			chiral (15, 8)    &   7.9002 & 1.558  & 1.6628 & -53/162 \\ 	
			armchair (20, 20) &  13.4365 & 2.659  & 2.8393 & -50/300 \\
			\hhline{=|=|=|=|=}
		\end{tabular}
	}
	\label{tab:Ex1}
\end{table}

\subsection{MD to FEM Comparison 2: Spatial Nanotruss}
\label{ex_3DMD_FEM_Truss}
\begin{figure}
	\centering
	\includegraphics[scale=1]{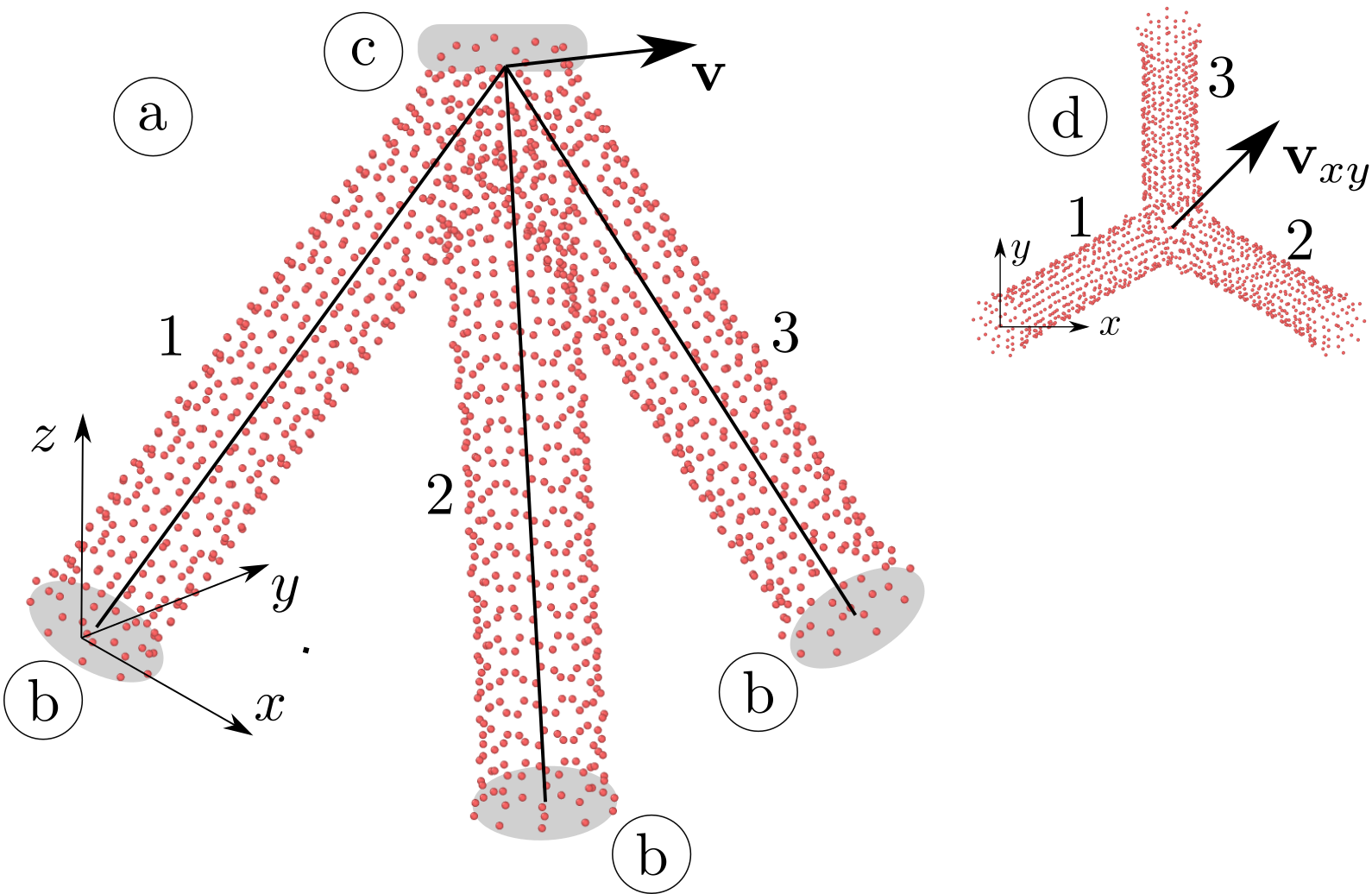}
	\caption{MD truss (a): geometry, coordinate system, hinged supports for CNTs 1-3 (b), velocity $\mathbf{v}=\left\langle 0.05, 0.05, 0.05 \right\rangle$ (c), and its projection $\mathbf{v}_{xy}$ to $xy$ plane (d).} 
	\label{fig:MDtruss}
\end{figure}

The purpose of the present example is to demonstrate the accuracy of the proposed finite element formulation. To this end, the loading of a simple three-dimensional nanotruss structure is considered, Fig.~\ref{fig:MDtruss}. The vertices of the structure are the vertices of a tetrahedron. The edges of the base are 4.899 nm long, while the edges connecting the vertices of the base to the top vertex are 4.923 nm long. In this way, the latter edges form a spatial nanotruss. The structure is loaded at the top vertex, while three hinged supports are placed at the other ends, Fig.~\ref{fig:MDtruss}. Each edge is (10, 0) SWCNT with a diameter of 0.7827 nm.

This structure is then analyzed with the newly developed NN-FEs and the results are compared with those obtained with MD. The generation of the FEM model is straightforward, and the prescribed displacements are used as loads. On the other hand, the MD model requires careful preparation to ensure proper modeling of the geometry and boundary conditions. The starting point is a (10, 0) SWCNT. The atoms of the SWCNT are rotated in space to obtain the required structure. The result is that where three SWCNTs are connected, the atoms of all three SWCNTs overlap. Such an arrangement of atoms would cause atoms to be ejected from the simulation box due to the high interatomic bonds between  closely positioned atoms. To solve this issue, pairs of atoms that are closer than 0.04 nm are identified and one of the overlapping atoms is deleted. Then, all the remaining atoms that are 4 nm or more away from the base ($xy$ plane or $z\ge4$ nm) of the tetrahedron form a rigid body that is used to introduce the loading. The final structure consisted of 1358 atoms. Subsequently, the nanotruss is equilibrated at 300 K, and during the equilibration process it expands and reaches the above dimensions.

In the MD simulation, the velocity of this rigid body is prescribed  to introduce a load. Three different but constant velocity vectors $\mathbf{v}$ (nm/ps) were considered:
\begin{enumerate}[label=Case \alph*:,leftmargin=4\parindent]
	\item $\mathbf{v}=\left\langle 0.05, 0.05, 0.05 \right\rangle$ 
	\item $\mathbf{v}=\left\langle 0, 0, 0.05 \right\rangle$ 
	\item $\mathbf{v}=\left\langle 0, 0, -0.05 \right\rangle$ 
\end{enumerate}
The time step was 0.001 ps. Thus, one case represents a general spatial deformation of the truss, while the other two are extension and compression, respectively. Further, the last rings of atoms at the other ends of the SWCNTs are modeled as rigid bodies that can rotate about the center of the ring while at the same time the center cannot translate. This effectively models hinged supports. All other settings are the same as in Sec.~\ref{sec_MD_model}. The analysis lasted until one or more atoms left the simulation box or up to \numprint{300000} steps (which corresponds to 300 ps).

\begin{figure}
	\centering
	\includegraphics[scale=1.2]{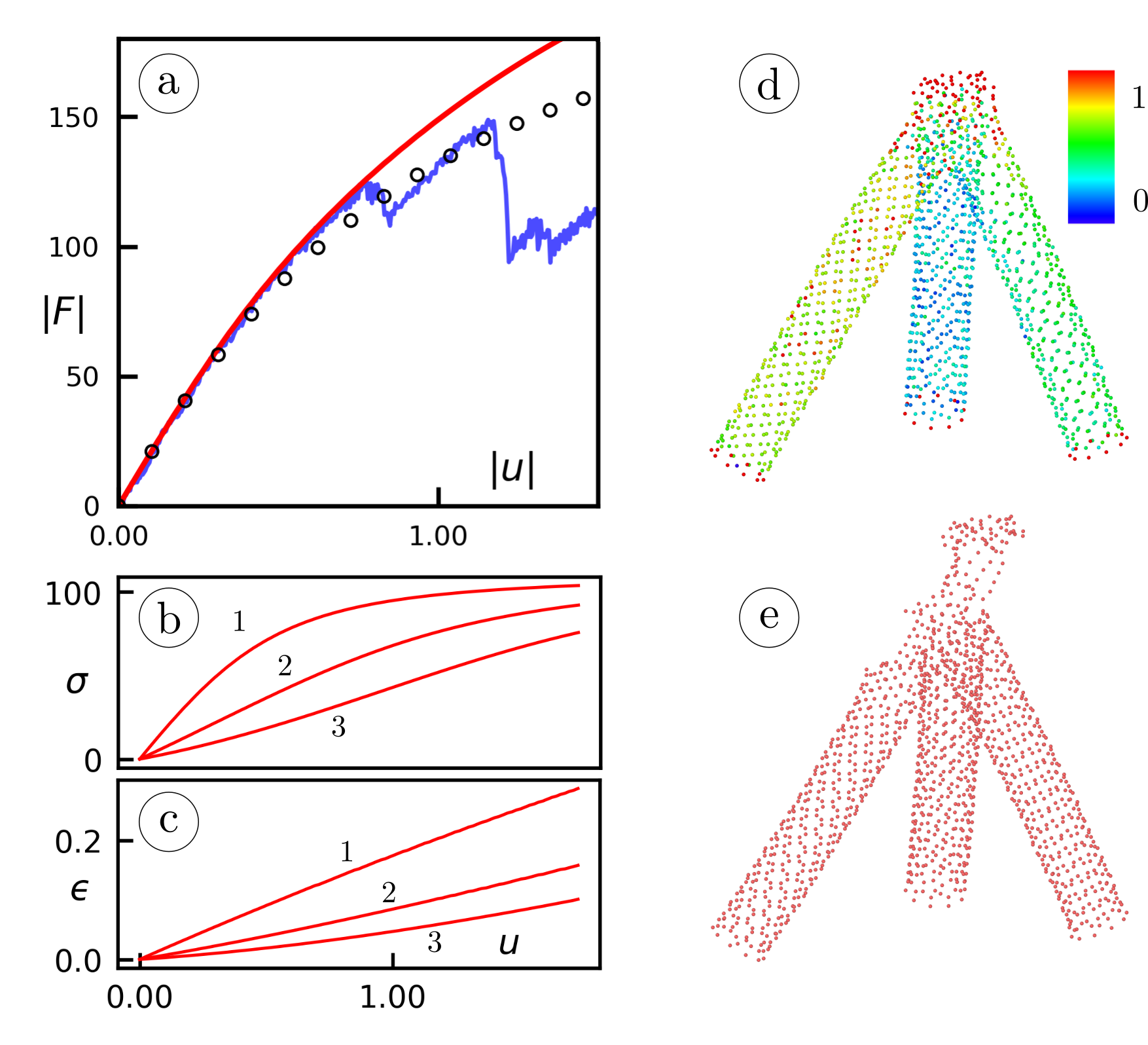}
	\caption{Results for velocity vector $\mathbf{v}=\left\langle 0.05, 0.05, 0.05 \right\rangle$: (a) force-displacement curve as obtained from MD (blue) and FEM (red). The black circles represent the results obtained with diameters calculated with the incompressible formulation as readily available in Abaqus. (b) axial stress - displacement in CNTs 1-3 (FEM), (c) axial strain - displacement in CNTs 1-3 (FEM), (d) distribution of virial von Mises stresses, (GPa nm$^3$), (e) failure mode.} 
	\label{fig:MDtruss_spatial_setup}
\end{figure}

\begin{figure}
	\centering
	\includegraphics[scale=1.2]{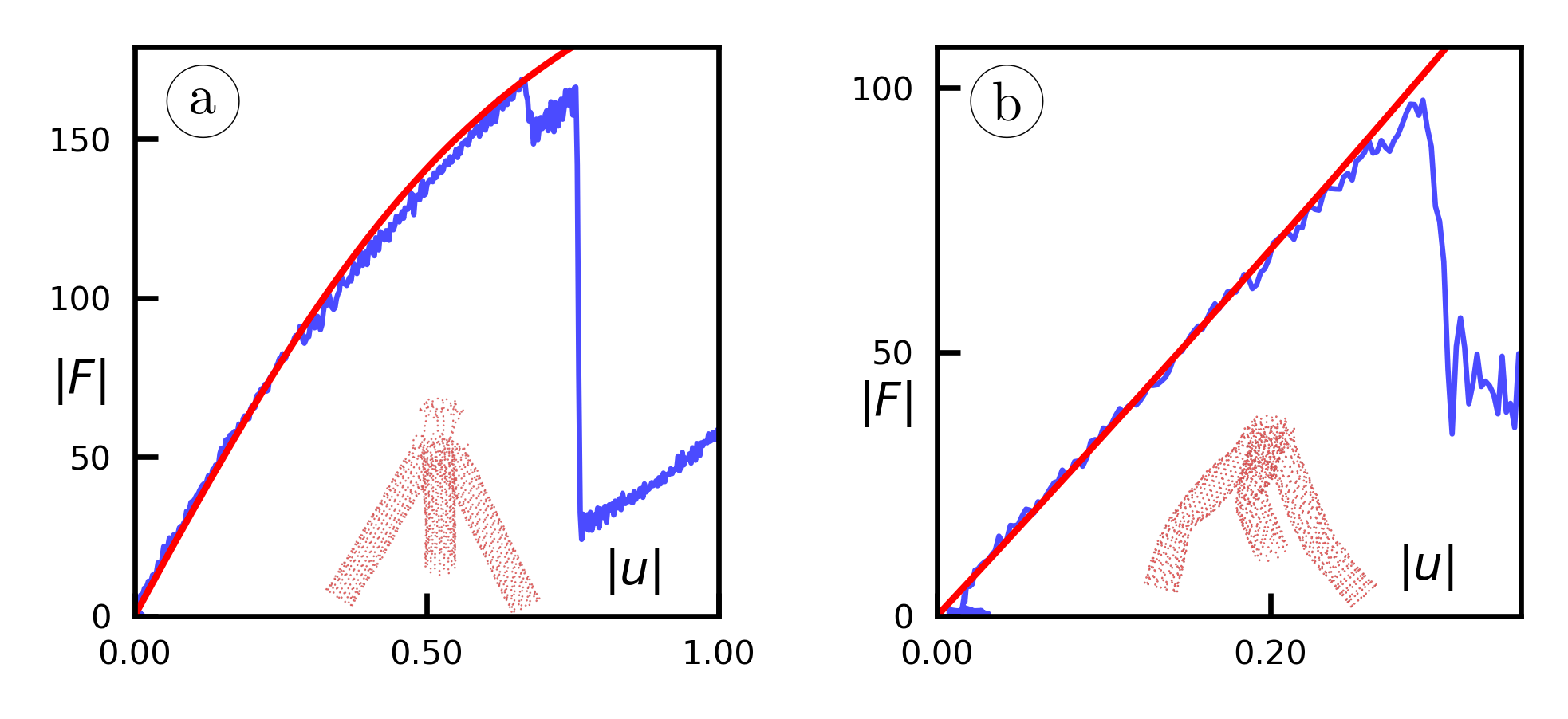}
	\caption{Force - displacement curves as obtained from MD (blue) and FEM (red) for velocity vectors $\mathbf{v}$ (a) $\left\langle 0, 0, 0.05 \right\rangle$ (b) $\left\langle 0, 0, -0.05 \right\rangle$. The insets show corresponding failure modes of the nanotruss.} 
	\label{fig:MDtruss_results_compress_extend}
\end{figure}

The results obtained with MD and FEM are shown in Figs.~\ref{fig:MDtruss_spatial_setup}-\ref{fig:MDtruss_results_compress_extend}. Perhaps the first thing that may be noticed is that FEM closely follows MD force-displacement curves, Fig.~\ref{fig:MDtruss_spatial_setup}a \& \ref{fig:MDtruss_results_compress_extend}. The discrepancies in Fig.~\ref{fig:MDtruss_results_compress_extend}a,b are due to the repositioning of the atoms. Such repositioning alters the SWCNT configuration locally, but sufficiently to cause partial unloading of at least one CNT, which is visible as a short sudden drop. Additionally, it changes the tangent modulus. The same can be noted in Fig.~\ref{fig:MDtruss_spatial_setup} upon closer examination. As visible in the insets, the structure fails near the rigid body used to introduce the load in spatial deformation and tensile deformation. For compressive deformation, Fig.~\ref{fig:MDtruss_results_compress_extend}b, a single CNT buckles first, while the other two buckle very soon after the first one. For the spatial deformation, the results obtained by the diameter calculated with the incompressible formulation are also given. While reasonable accuracy is obtained for smaller displacements, larger displacements result in a loss of accuracy when such an approach is followed.

The evolution of true axial stress and true strain during spatial deformation as obtained from FEM is provided in Fig.~\ref{fig:MDtruss_spatial_setup}b, c. The distribution of virial von Mises stresses is given in Fig.~\ref{fig:MDtruss_spatial_setup}d. The virial stresses must be divided by the corresponding volume to obtain the usual stresses, and no attempt is made to relate them to the stresses obtained by FEM in Fig.~\ref{fig:MDtruss_spatial_setup}b. Nevertheless, the nonhomogeneous distribution of stresses is clearly evident. The highest stresses occur near the connection of three CNTs, which is clearly a trigger for the fracture of the CNT designated as No. 1 in Fig.~\ref{fig:MDtruss}. Similar behavior is observed during extension, where the truss again fails at the joint of the rigid and flexible parts of the truss, Fig.~\ref{fig:MDtruss_results_compress_extend}a.


\subsection{Nonlinear Static Analysis of a Three-Dimensional Nanotruss}
\label{ex_3DTruss}
\begin{figure}
	\centering
	\includegraphics[scale=0.3]{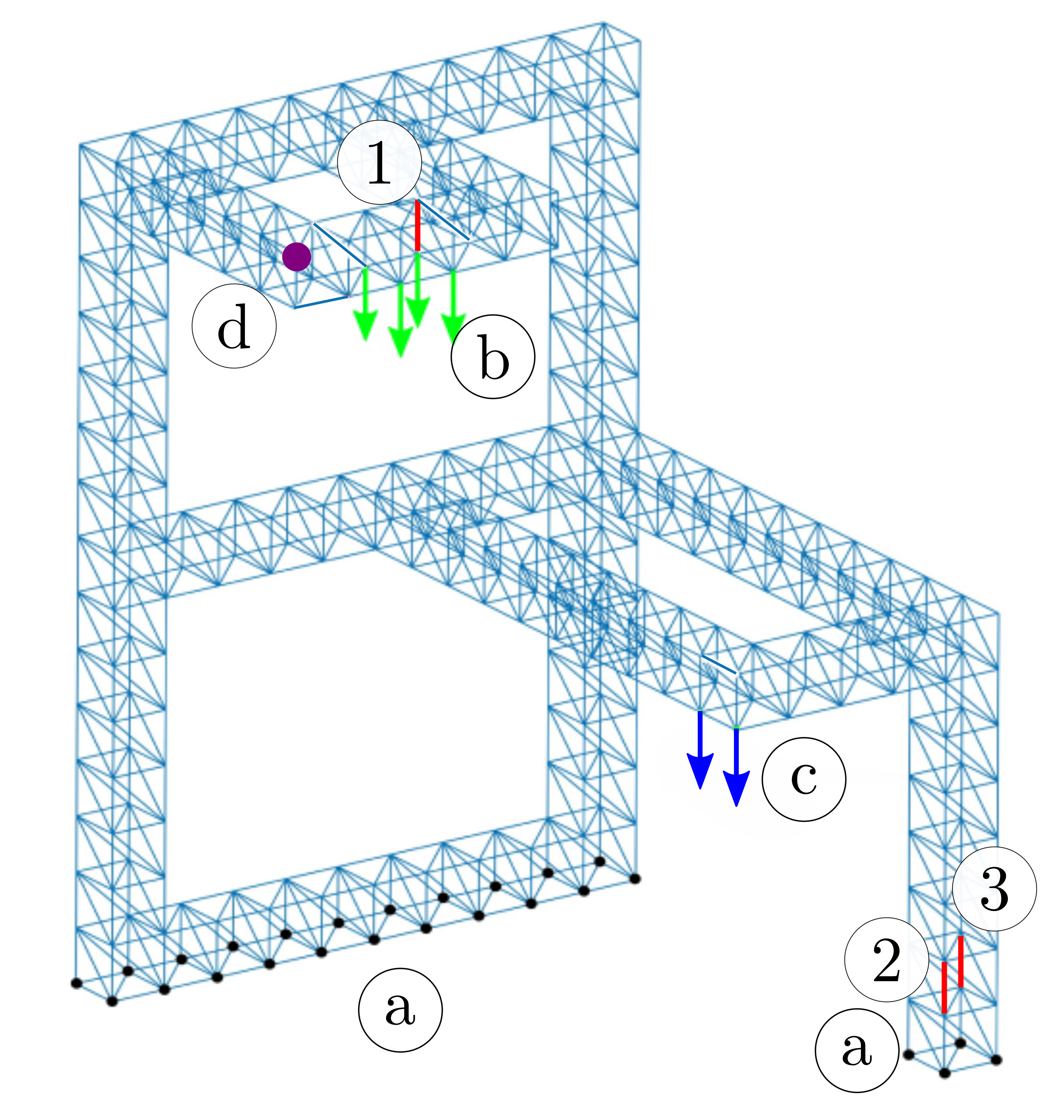}
	\caption{3D truss: geometry, boundary conditions and locations where results are reported. (a) supports, (b) prescribed displacement, (c) forces, (d) position of maximum displacement, (1-3) elements where stress-strain curves are reported.} 
	\label{fig:3dtruss}
\end{figure}

The present example aims at testing the robustness and convergence properties of the new framework on a medium-sized problem. The truss topology of the present example stems from \citep{eggersmann2019model}, where it was used in a different context. The supports remain the same, while a different loading is assumed here. The structure is assembled of basic building elements - cubes formed by rods, Fig.~\ref{fig:3dtruss}. Each edge of the cube is 10 nm long and made of (7, 7) armchair SWCNTs with a cross-sectional area of 0.9999 nm$^2$. The overall dimensions of the truss are $150\times100\times110$ nm. In total, the nanotruss is comprised of \numprint{1246} truss elements and 376 nodes.

The translational degrees of freedom at all nodes at the bottom of the truss are removed to support the truss (indicated by (a) in Fig.~\ref{fig:3dtruss}). The loading is represented by two sets of loads - prescribed displacements (b) and two concentrated forces (c). Both loads are applied incrementally, increasing by 2 nm and 1 nN in each increment.

\begin{figure}
	\centering
	\includegraphics[scale=0.7]{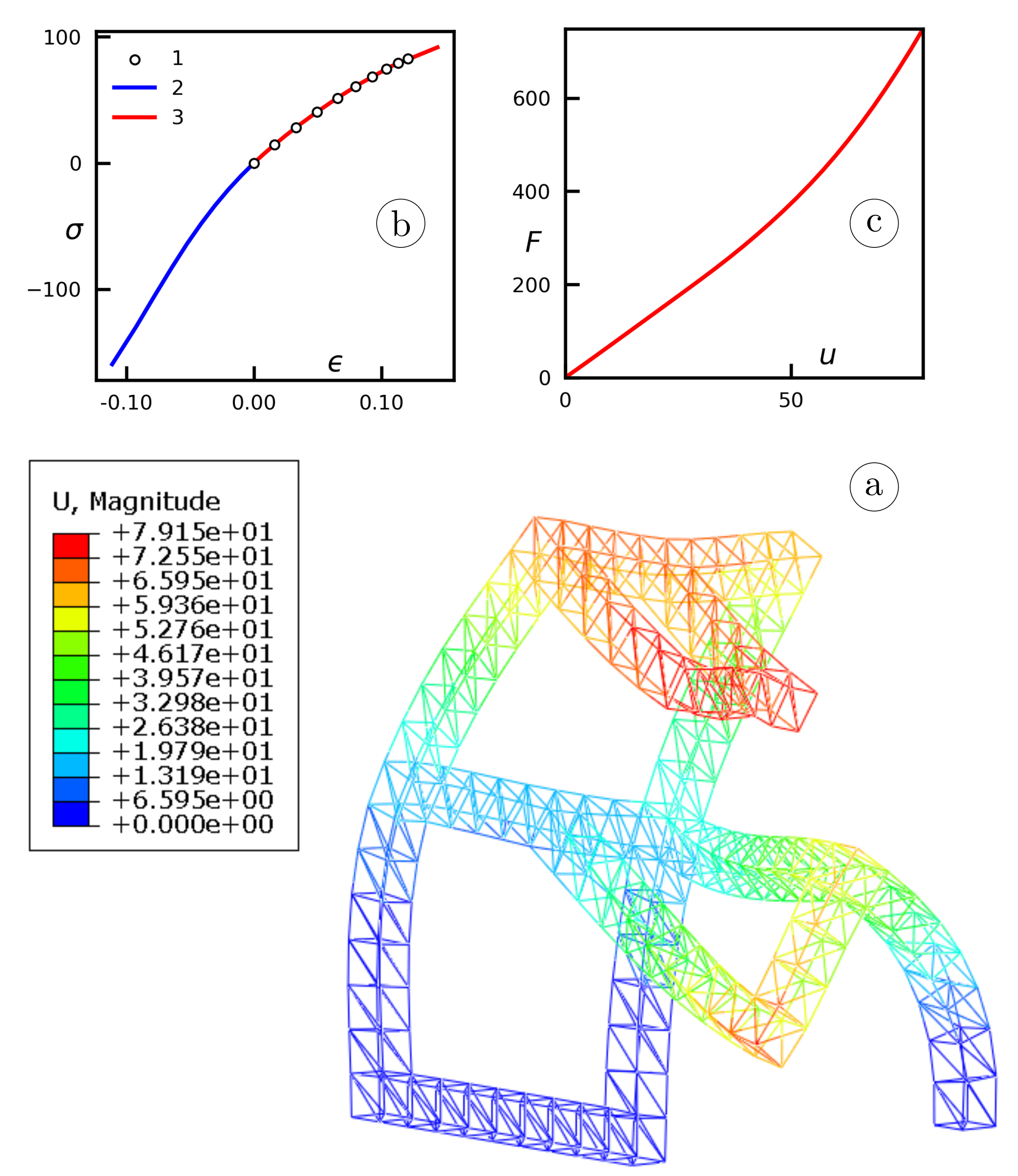}
	\caption{(a) Deformed shape of the truss at the end of simulation (true scale), (b) stress-strain curves for elements 1-3, (c) total support reaction vs. displacement curve.} 
	\label{fig:3dtruss_res}
\end{figure}

The analysis fails after 28 increments and 58 iterations, so the average number of iterations per increment is 2.07. Thus, the convergence can be judged to be quite fast. The maximum values of loads at failure were 56 nm or 28 nN in each loaded node. The final results are given in Fig.~\ref{fig:3dtruss_res}. The maximum displacement was 79.15 nm in the node marked with purple color (d), while the minimum and maximum stresses were -170.9 GPa and 97.0 GPa, respectively. The stress-strain curves are given for the elements where the minimum (2) and maximum (3) stresses are observed, as well as for the element (1) located near the node with the maximum displacement. These elements are marked in red in Fig.~\ref{fig:3dtruss}. For further illustration, the change in displacement in a node where the maximum is observed is plotted against the sum of all support reactions.

As for the robustness of the formulation, it can be partially estimated by observing the behavior for data that are outside the dataset range on which NN was trained. For the (7, 7) SWCNT, minimum/maximum stress and strain of -52.35/102.40 GPa and -0.04450/0.1879, respectively, were obtained using MD. It can be judged that the tensile stresses in the present case are still lower than the maximum stresses observed in uniaxial tensile MD simulations of (7, 7) SWCNT. However, the compressive stresses are more than three times higher than those observed in the MD simulation of uniaxial compression. Nevertheless, at least some dataset compressive stresses were larger than the stress at failure, albeit for different SWCNT configurations than those considered in the present example (the lowest stress of -178.29 GPa and the lowest strain of -0.156 in the dataset for all SWCNT configurations were observed for the (5, 0) nanotube). This means that the NN is trained with stresses and strains of these levels, but for different chiralities. It should also be recalled that the SWCNTs with the smallest diameter have significantly larger maximum compressive stresses than the other configurations, resulting in steep gradients in the NN approximation. Another factor affecting failure is large geometric nonlinearities, where the maximum total displacement is greater than half the height of the structure. Overall, the performance can nevertheless be judged to be quite favorable.

\subsection{Metamaterials Design: Size Effects}
\label{ex_sizeffects}
\begin{figure}
	\centering
	\includegraphics[scale=1.1]{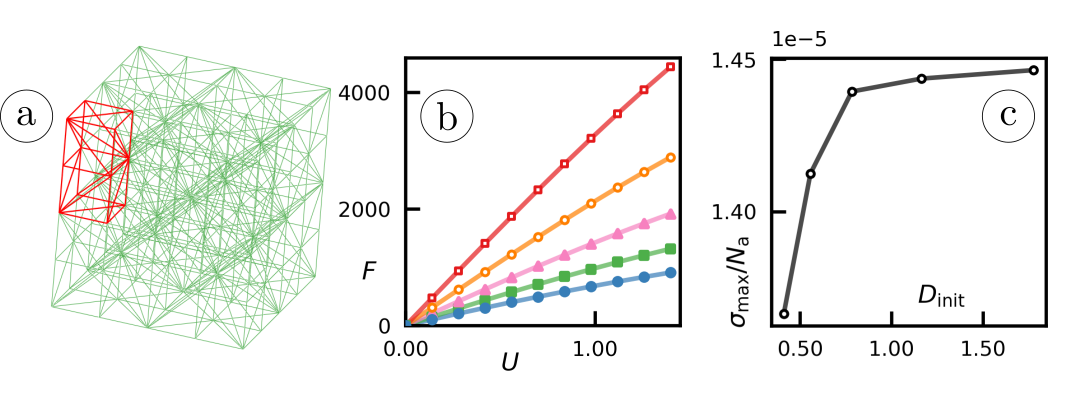}
	\caption{Nanotruss used to estimate size effects: (a) model, red lines indicate the basic element, (b) force (nN)-displacement (nm) curves for (5, 0) (bottom), (7, 0), (10, 0), (15, 0), and (23, 0) (top) SWCNTs (c) maximum averaged tensile stress in RVE $\sigma_\text{max}=F_\text{max}/A$ per atom (GPa/atom) vs. initial diameter (nm).} 
	\label{fig:size_effects}
\end{figure}

Alternatively to the size effects typically studied in a single SWCNT, a nanotruss structure built from SWCNTs is used as a building block of a hypothetical material and studied for the influence of SWCNT diameter, Fig.~\ref{fig:size_effects}a. The basic part used to build the representative volume element (RVE) is highlighted in red. It is mirrored and rotated to obtain a symmetric structure. In total, there are 125 nodes and 574 T3D2 finite elements. Finite element properties are assumed to resemble 5 different zigzag SWCNTs: $(5, 0)$, (7, 0), (10, 0), (15, 0) and (23, 0), with initial diameters listed in Tab.~\ref{tab:Ex2_size}. The structure is loaded in tension by the prescribed displacement of 1.4 nm on one side. The supports are located on the opposite side and allow free contraction of the structure. The overall dimensions of such a representative volume element are $20 \times 20 \times 20$ nm.

The size effects are expected to occur due to the different diameters of the CNTs involved. As the cross-sectional area of each truss element increases with diameter, the structure becomes stiffer. In other words, the imposed load is distributed over a larger number of atoms. Since both the topology and the volume are kept constant, the number of atoms also increases as the diameter increases, and eventually the mass density, Tab.~\ref{tab:Ex2_size}, also increases. Therefore, it would also be interesting to see if the stress divided by the number of atoms (or alternatively by the mass density) is affected by the size effects.

Selected results are given in Tab.~\ref{tab:Ex2_size} and Fig.~\ref{fig:size_effects}b, c. The mass density approaches rather quickly the usual density of carbon of \numprint{1810} (amorphous state) - \numprint{3515} (diamond) kg/m$^3$. Tab.~\ref{tab:Ex2_size} also lists the maximum and minimum stresses in CNTs at the end of the simulations. Maximum stresses occur in CNTs that are parallel to the loading direction, while the minimum ones occur in planes that are orthogonal to the loading direction. A clear trend of increase in $\sigma_\text{CNT,max}$ can be seen, converging to a certain limit. The compressive stresses $\sigma_\text{CNT,min}$ show some oscillations.

To gain insight into the overall behavior of the RVE, the curves of total force $F$ vs. displacement $U$ are plotted in Fig.~\ref{fig:size_effects}b. A clear increasing trend is visible as expected, and convergence to a particular limit curve is not observed. The average engineering axial stress in the structure can be evaluated as $\sigma=F/A$, where the cross-sectional area is $A=20 \times 20$ nm$^2$. Plotted against the engineering strain, these would naturally follow the same trend. To obtain the influence of the number of atoms mentioned above, these average stresses can be divided by the total number of atoms in the structure $N_\mathrm{a}$. Unlike the $F-U$ curves, these results do converge to a limit point. This becomes even clearer if only the maximum stresses $\sigma_\text{max}=F_\text{max}/A$ are plotted against the initial diameter, Fig.~\ref{fig:size_effects}c. Thus, the average stresses increase as the initial CNT diameter increases, but ratios of these to number of atoms or the mass density no longer increases, giving a kind of size dependence.

At the end, it is emphasized that failure behavior is not analyzed. It is only briefly noted that the compressive stresses $\sigma_\text{CNT,min}$ will be the cause of failure, with the largest diameter CNTs being the first to fail by shell-like buckling. We also ignore that for (5, 0) and (7, 0) the condition $L/D < 12.5$ is violated, so they could fail by beam-like buckling. 

\begin{table}
	\caption{Data about CNTs and RVEs in Ex. \ref{ex_sizeffects}}
	\centering
	{		\footnotesize
		\begin{tabular}{c|c|c|c|c|c}
			\hhline{=|=|=|=|=|=}
			CNT 			 &  $N_\text{a}$ &  $D_\text{init}$ & $\rho_\text{RVE}$ & $\sigma_\text{CNT,min}$ & $\sigma_\text{CNT,max}$ \\
			configuration 	 &    &  (nm)  & (kg/m$^3$) & (GPa) & (GPa) \\	
			\hline
			(5, 0)  &  \numprint{166890} & 0.411  & 416 & -24.38 & 54.34 \\ 
			(7, 0)  &  \numprint{233646} & 0.556  & 583 & -25.32 & 58.72 \\ 	
			(10, 0) &  \numprint{333780} & 0.783  & 832 & -25.43 & 61.49 \\
			(15, 0) &  \numprint{500670} & 1.163  & \numprint{1248} & -25.21 & 62.83\\
			(23, 0) &  \numprint{767694} & 1.777  & \numprint{1914} & -25.14 & 63.48 \\
			\hhline{=|=|=|=|=|=}
		\end{tabular}
	}
	\label{tab:Ex2_size}
\end{table}

\subsection{Metamaterials Design: Fractal Octahedron Lattice}
\label{ex_LargeScale}
\begin{figure}
	\centering
	\includegraphics[scale=0.4]{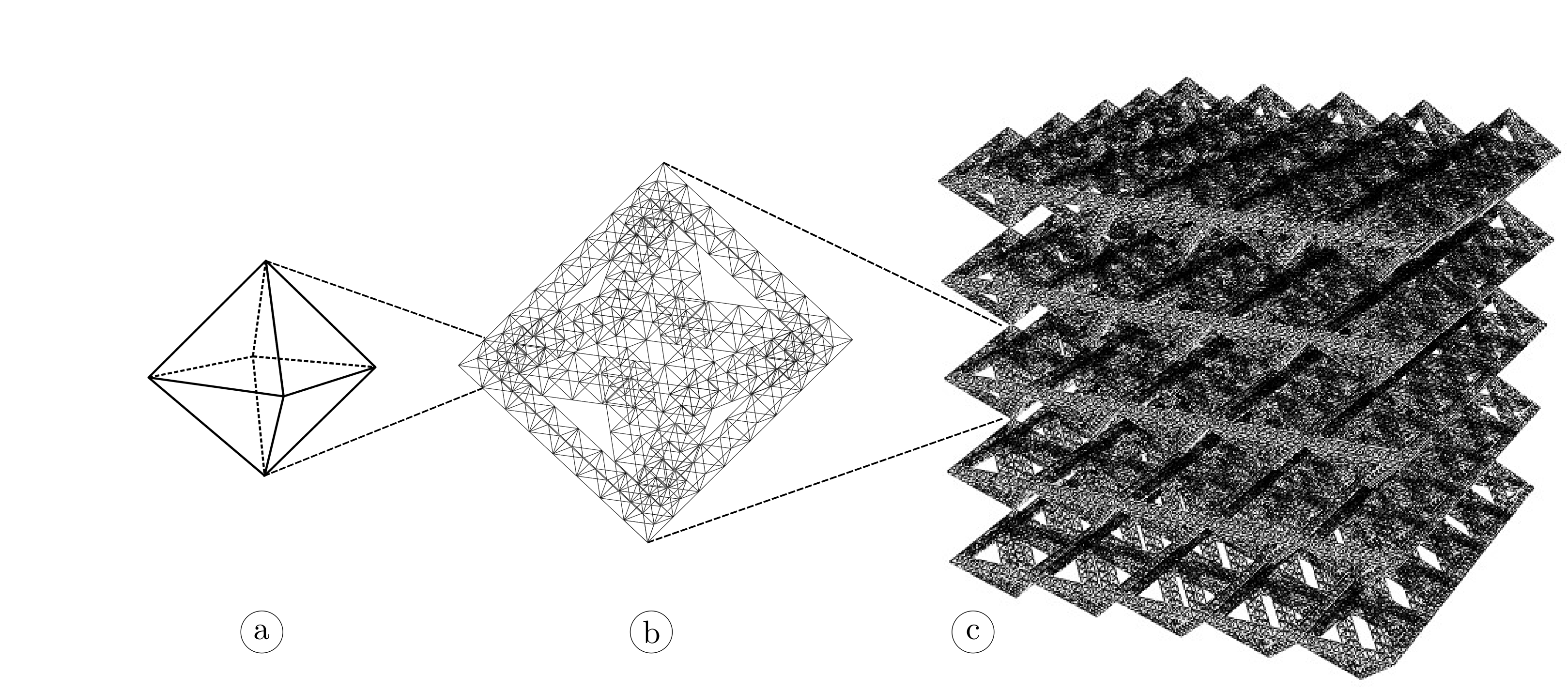}
	\caption{Topology of the nanolattice. (a) First-order octahedron, (b) second-order octahedron of octahedra, (c) nanolattice.} 
	\label{fig:octahedrons}
\end{figure}

Nowadays, carbon nanotubes and graphene are used to produce ultralight materials known as carbon aerogels \citep{sun2013multifunctional}. Although these materials are very light, the addition of carbon nanotubes allows them to withstand very large loads, considering the mass of the structure. In the latter work, an aerogel with a mass density of 1.15 mg cm$^{-3}$, a tensile strength of 10.9 kPa, and a Young's modulus of 329 kPa at room temperature is reported. Motivated by these results, we want to demonstrate the applicability of the present framework to an engineered structure of hypothetical lightweight metamaterials (not necessarily ultralight). In this example, the metamaterial is based on a higher-order octahedron of octahedra with dimensions $0.4\times0.4\times0.4$ \textmu m. As shown in Fig. \ref{fig:octahedrons}, the initial first-order octahedron is a repeated unit cell for the second-order octahedron of octahedra, which are then arranged in a nanolattice. This fractal-like geometry was inspired by the hierarchical 3D nanolattices described in \cite{meza2015resilient}. In contrast to the latter reference, the zeroth-order element used in the present nanostructure is an armchair (7, 7) CNT.

The architecture of the structure can be described as follows. Each first-order octahedron consists of two pyramids with (7, 7) SWCNTs as edges. Each first-order octahedron consists of 4 edges that are 10 nm long and 8 edges that are 8.66 nm long. Overall, the nanostructure is composed of \numprint{116900} elements, i.e., there are \numprint{37900} shorter and \numprint{79000} longer SWCNTs. The total length of all SWCNTs is then \numprint{1063160} nm. Since (7, 7) SWCNTs have an average of 119.6 atoms per 1 nm, the total number of atoms in the nanolattice is estimated to be \numprint{127151456}. A model with such a large number of atoms is a challenging task for any molecular dynamics code available today. The proposed approach is considerably faster. To ensure the validity of the results in the compressive regime, the aspect ratio must be below the transition point to beam-like buckling at $L/D < 12.5$, see Sec.~\ref{sec_MD_model}. Since the longest nanotube is 10 nm long and the diameter is 0.927 nm, the aspect ratio is within these limits and SWCNTs fail by shell-like buckling.

SWCNT stress and geometry data are given in Ex.~\ref{ex_3DTruss}. The boundary conditions resemble a uniaxial tension/compression test performed with a prescribed displacement of 0.1 \textmu m in all nodes at the octahedron vertices on one side of the lattice, while the supports on the opposite side allow unrestricted expansion and contraction. Note that for such a loading case a smaller model may suffice, for instance, a second-order lattice. However, to demonstrate that the present approach can efficiently handle very large nanostructures, the considered model is chosen to be larger than necessary.

\begin{figure}
	\centering
	\includegraphics[scale=1.0]{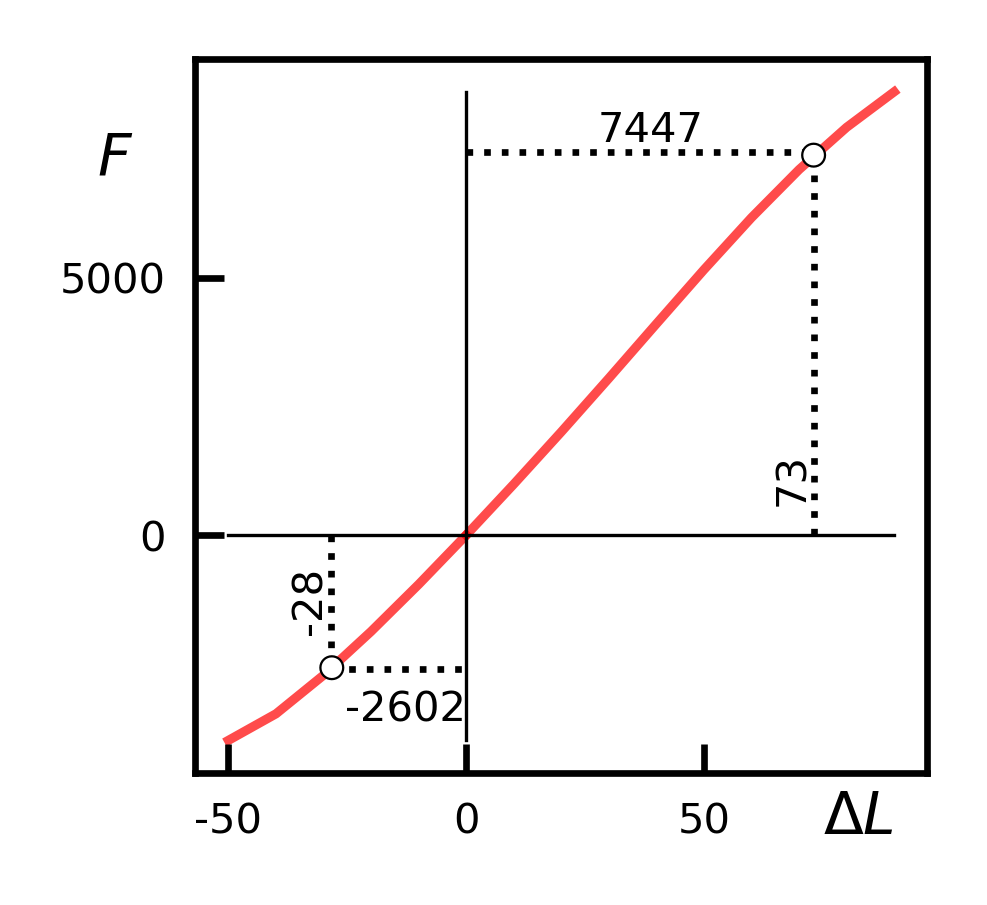}
	\caption{Force (nN)-displacement (nm) curves in tension and compression. Values at dotted lines correspond to points where the failure of the first SWCNT take place, as obtained by MD simulations for (7, 7) SWCNT.} 
	\label{fig:octahedron_FDL}
\end{figure}

In both the tension and compression tests, the analysis is performed in 10 equal increments. The analysis fails in the tensile test in the 10$^\text{th}$ increment. The total number of iterations in 9 converged increments was 29. Using the linear approximation between two increments, the ultimate stresses obtained in the uniaxial MD test are reached at a displacement of 73 nm, with a corresponding tensile force of 7447 nN, Fig.~\ref{fig:octahedron_FDL}. After this point, the analysis continues until it fails in the 10$^\text{th}$ increment as mentioned above. For compression, the analysis fails in the 5$^\text{th}$ increment, while the ultimate stresses are reached at $\Delta L=-28$ nm and $F=-2602$ nN. It is noteworthy that both curves are nearly linear in both tensile and compressive tests. Nonlinearities appear only when the dataset's upper and lower limits of (7, 7) SWCNT are exceeded.

\begin{figure}
	\centering
	\includegraphics[scale=0.55]{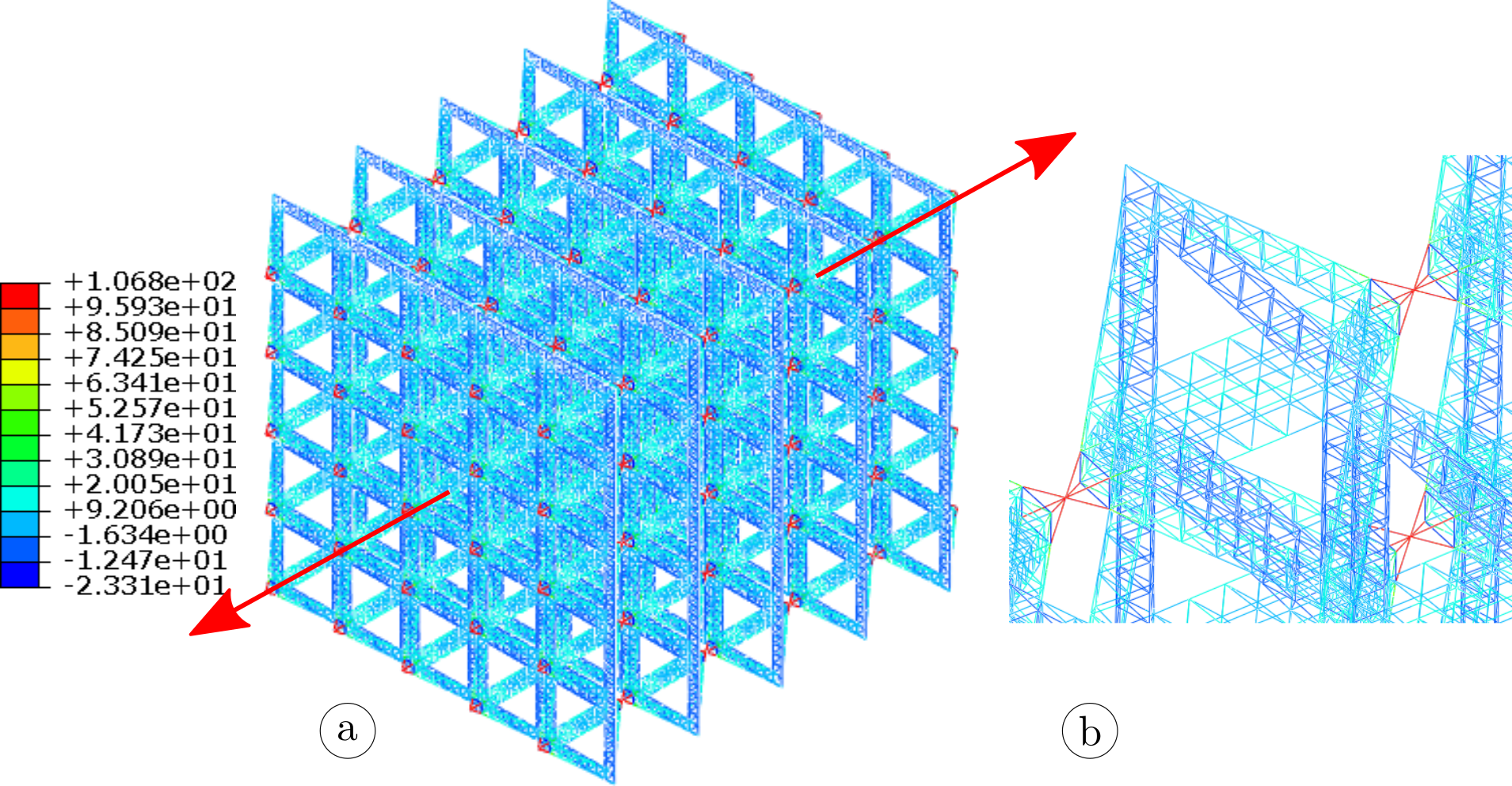}
	\caption{Stresses in tension at $\Delta L=80$ nm; (a) full model, (b) detail. Red arrows indicate the direction in which loading is applied.} 
	\label{fig:octahedron_stress_zoom}
\end{figure}

\begin{figure}
	\centering
	\includegraphics[scale=0.55]{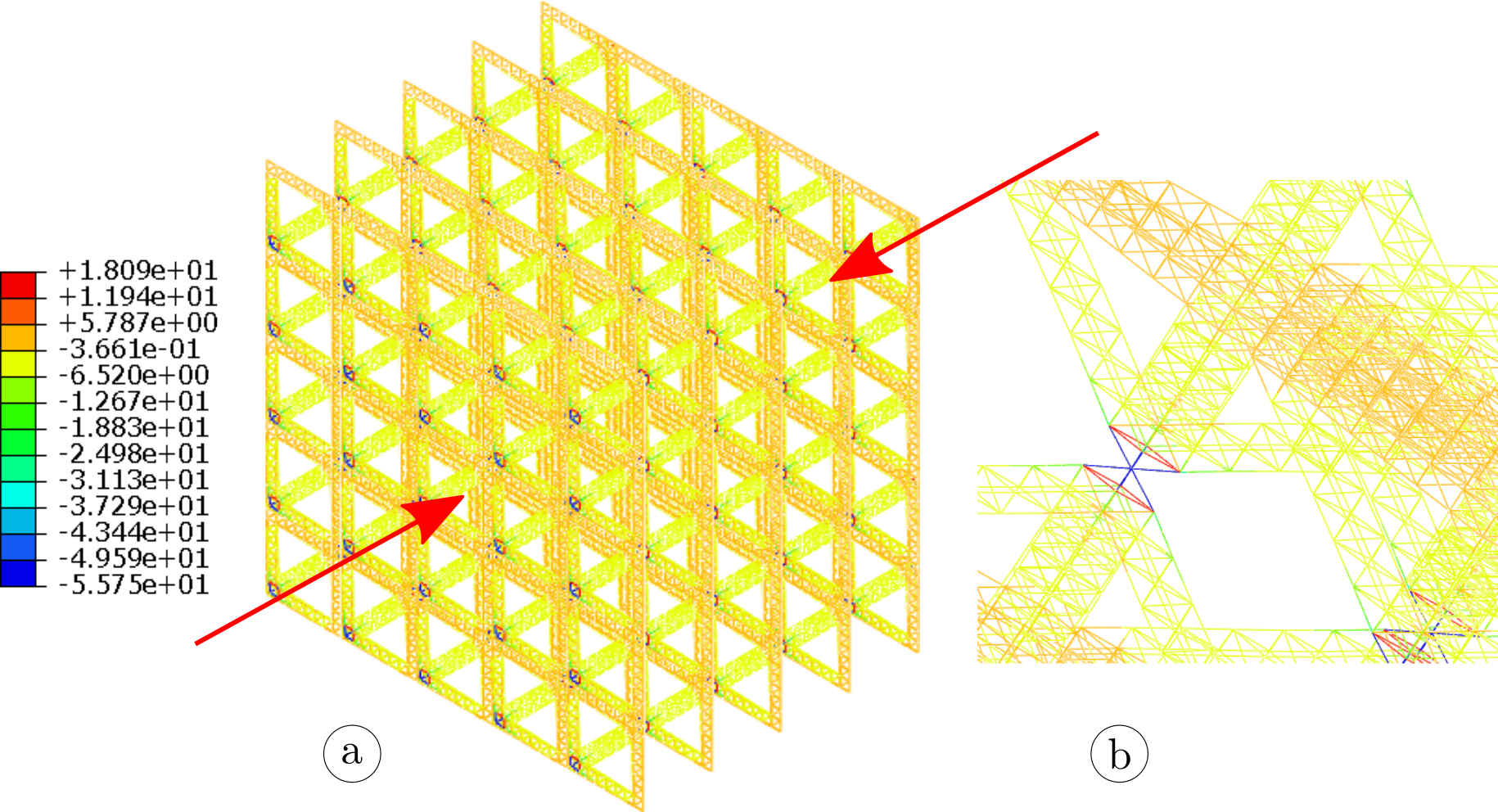}
	\caption{Stresses in compression at $\Delta L=30$ nm; (a) full model, (b) detail.} 
	\label{fig:octahedron_stress_zoom_comp}
\end{figure}

Figs. \ref{fig:octahedron_stress_zoom}-\ref{fig:octahedron_stress_zoom_comp} present axial stresses in CNTs during tests. The largest tensile stresses occur at the joints connecting two second-order octahedron of octahedra, while the largest compressive stresses occur at the bases of second-order octahedron of octahedra. The stresses are lowest at the edges of the lateral faces of second-order octahedron of octahedra.

At the end, having in mind the above number of carbon atoms in the structure and the volume of the structure, the mass density can be evaluated as 39.62 kg/m$^3$. If the tensile force at fracture 7447 nN is divided by the area $A=0.4\cdot0.4$ \textmu m$^2$, then the average engineering axial stress at the first CNT fracture in the nanostructure is 46.5 MPa, which can be alternatively expressed per mass density as 1.174 MPa m$^3$/kg. Likewise, using the engineering stress and strain, the linear approximation in the first increment gives a Young's modulus of 247 MPa. As in the compression test, the average axial stress at failure is -16.3 MPa or -0.410 MPa m$^3$/kg.

\section{Closing remarks}
The present investigation has shown that it is possible to obtain results that can reproduce the uniaxial mechanical behavior of carbon nanotubes fairly accurately by coupling MD-ML-FE methods. Although molecular dynamics simulations still have to be performed to establish an initial NN model, having a computationally affordable surrogate can be of great benefit. The number of constraints regarding the application of the classical FEM approach to nanotrusses is significantly reduced. The main findings are recapitulated once more:
\begin{itemize}
	\item The proposed approach successfully circumvents the need for an explicit introduction of the nonlocal parameter, as the NN assumes the role of the constitutive model.
	\item Uniaxial compression MD simulations were performed, and when these data were added to the previously reported tensile tests, the most comprehensive MD dataset on SWCNT uniaxial behavior in existence was obtained. This provided clear buckling limits for all configurations considered. It is confirmed that shorter SWCNTs fail due to shell-like buckling. In addition, it is noted that when ML training is performed on both tension and compression data, the Young's modulus distribution among SWCNT configurations at the zero strain state is different than when only tension data is considered, as is usually the case.
	\item Using such a dataset, the PICINN model is developed that can be used as a surrogate for classical constitutive models, as long as the boundary conditions used in the uniaxial MD tests are respected. Unlike existing constitutive models used in FEM, which can be applied to specifically selected SWCNTs in a particular regime, typically linear, this model covers all possible SWCNT configurations up to 4 nm at all strains. Since the dependence on length is practically absent in shell-like buckling of thin cylinders, the NN model is valid in compression at least until the beam-like buckling limit is reached at aspect ratio $L/D=12.5$.
	\item MD simulations described in the literature are usually performed at 0 K. The analysis of real problems requires the introduction of temperature, which inevitably involves stochastic vibrations of SWCNT atoms. This is a significant complication. The present framework solves the problem by relying on the recently proposed partially input convex integrable neural networks, which provide a physically sound convex constitutive model. Improving PICINN in the form of weight normalization shortens the training process and provides better accuracy. ML framework also enforces $\sigma=0$ and $\psi=0$ at $\epsilon=0$. Note that an alternative approach could be based on \cite{thakolkaran2022nn}, i.e., relying on physically informed NNs trained on displacements and support reactions, but this remains to be explored.
	\item The PICINN is implemented in Abaqus by coupling the capabilities of user subroutines with the TensorFlow-Keras code. This enables very practical applications of truss finite elements to nanoscale structures. The flexibility of Abaqus features, including handling geometric nonlinearities, provides a very general finite element framework.
	\item The Abaqus implementation also required additional extension for larger strain regimes. In particular, for truss elements, the Abaqus code assumes incompressible behavior when geometric nonlinearities are involved. This leads to discrepancies with the results from MD due to miscalculations of the cross-sectional area. The problem is solved by training an additional NN that provides SWCNT diameter as a function of chirality and strain. The new NN is then used to rescale the cross-section and provides the correct cross-sectional area. The extension is not required for small strain analysis.
	\item Finally, the Examples section provides a comprehensive set of benchmarks. Starting with the simplest example - reproducing MD uniaxial tests by the FEM framework shows excellent accuracy. The justification of the proposed approach is further confirmed by comparing the results of MD and FEM in three different loading cases of a simple spatial nanotruss. The robustness is then demonstrated on a medium-sized nanotruss. The framework opens up possibilities for the design of metamaterials that would otherwise require very computationally intensive MD procedures. This is illustrated with an analysis of size-effects and on a very large fractal octahedron lattice.
\end{itemize}

\section*{Data availability}
Datasets, trained models and numerical MD data from Ex.~\ref{ex_3DMD_FEM_Truss} are available at 

 \href{http://dx.doi.org/10.17632/k7twggfsdm.1}{http://dx.doi.org/10.17632/k7twggfsdm.1}.

\section*{Acknowledgments}
This work has been supported by the Croatian Science Foundation under the project IP-2019-04-4703, and under the project Young Researchers' Career Development Project - Training New Doctoral Students. This support is gratefully acknowledged.

\appendix
\section*{Appendices}

\section{Convexity Issues}
\label{sec:app1}
To ensure that the stress-strain curves increase monotonically, the starting point is the definition of stresses as $\sigma=\partial_\epsilon \psi(\epsilon,n,m)$, as obtained from the second law of thermodynamics. Monotonicity is assured if the strain energy $\psi$ is a convex function in $\epsilon$, which requires that the second derivative $\partial^2_{\epsilon\epsilon} \psi \ge 0$ is positive. At the same time, no convexity constraint is imposed with respect to the chirality parameters $n,m$. In this sense, the function $\psi$ can be represented by the PICINN \citep{amos2017input,huang2022variational}. The following proof represents a somewhat extended version of the proof in \cite{huang2022variational}.

pWN in Fig.~\ref{fig:NN} is considered here, but the procedure is the same for other configurations listed in Tab.~\ref{tab:NNs}. The input layer variables are denoted as $\mathbf{x}_0, \mathbf{y}_0$, where $\mathbf{x}_0=\left[ \epsilon \right] $ is the vector containing the variables where the function is convex, while $\mathbf{y}_0=\left[ n \quad m \right]^\mathrm{T}$ are the variables without convexity constraint. The output of NN is the strain energy $\mathbf{z}=\left[ \psi \right]$. In the PICINN, each hidden layer is assembled of two layers $\mathbf{x}_i, \mathbf{y}_i $, which are connected in some way to the next hidden layer $\mathbf{x}_j, \mathbf{y}_j$, where $j=i+1$. For the present case, these two layers are related as:
\begin{equation}
	\label{eq_app1_f}
	\begin{array}{l}
		\mathbf{x}_{j}=g \left\lbrace  \widetilde{\mathbf{W}}^{x_j}_{x_i y_i}\left[\mathbf{x}_i \odot g (\mathbf{W}_{x_i y_i} \mathbf{y}_i+\mathbf{b}_{x_i y_i})\right] \right.\\
		\quad \quad \left.+ \overline{\mathbf{W}}^{x_j}_{x_0 y_i}\left[\mathbf{x}_0 \odot (\mathbf{W}_{x_0 y_i} \mathbf{y}_i+\mathbf{b}_{x_0 y_i})\right] + \overline{\mathbf{W}}_{y_i} \mathbf{y}_i +\mathbf{b}_{y_i}  \right\rbrace \\
		\mathbf{y}_{j}=g(\overline{\mathbf{W}}_{y_i y_j}\mathbf{y}_i+\mathbf{b}_{y_i y_j}), 
	\end{array}
\end{equation}
where $\odot$ denotes the Hadamard product, $g$ is the activation function, $\widetilde{\mathbf{W}}^{x_j}_{x_i y_i}$, $\mathbf{W}_{x_i y_i}$, $\overline{\mathbf{W}}^{x_j}_{x_0 y_i}$, $\mathbf{W}_{x_0 y_i}$, $\overline{\mathbf{W}}_{y_i}$, $\overline{\mathbf{W}}_{y_i y_j}$ are weights, and $\mathbf{b}_{x_i y_i}, \mathbf{b}_{x_0 y_i}, \mathbf{b}_{y_i}, \mathbf{b}_{y_i y_j}$ are biases. In this sense, the above relations between layers can be viewed as compositions of functions and show the recursive nature of the NN.

With the relations between the layers defined, we would like to show that the strain energy $\psi$ calculated by the NN is indeed a convex function $\partial^2_{\epsilon\epsilon} \psi \ge 0$, i.e. $\partial^2_{\mathbf{x}_0\mathbf{x}_0} \mathbf{z} \ge 0$. For a start, it is demonstrated that the output of the $j$-th layer is convex with respect to the NN input $\partial^2_{\mathbf{x}_0\mathbf{x}_0} \mathbf{x}_j \ge 0$. The first and second derivatives can be evaluated as
\begin{equation}
	\label{eq_app1_x0_2nd}
	\begin{aligned}
		\partial_{\mathbf{x}_0} \mathbf{x}_{j} &=  \partial_{\mathbf{x}_{i}} \mathbf{x}_{j} \partial_{\mathbf{x}_0} \mathbf{x}_{i} \\
		\partial^2_{\mathbf{x}_0\mathbf{x}_0} \mathbf{x}_{j}&=\partial^2_{\mathbf{x}_{i}\mathbf{x}_{i}} \mathbf{x}_{j} 	(\partial_{\mathbf{x}_0} \mathbf{x}_{i})^2 + \partial_{\mathbf{x}_{i}} \mathbf{x}_{j} \partial^2_{\mathbf{x}_0\mathbf{x}_0} \mathbf{x}_{i}.
	\end{aligned}
\end{equation}
Consequently, noting that the sum of non-negative convex functions is a convex function \citep{boyd2004convex}, convexity is guaranteed if:
\begin{enumerate}[leftmargin=4\parindent]
	\item ${\partial^2_{\mathbf{x}_i\mathbf{x}_i} \mathbf{x}_{j}\ge0}$,
	\item $\partial_{\mathbf{x}_{i}} \mathbf{x}_j \ge 0$,
	\item ${\partial^2_{\mathbf{x}_0\mathbf{x}_0} \mathbf{x}_{i}\ge0}$.
\end{enumerate}
The first and third conditions can be interpreted as requiring that $\mathbf{x}_{j}$ be convex in $\mathbf{x}_i$ and $\mathbf{x}_{i}$ be convex in $\mathbf{x}_0$, while the second condition requires that $\mathbf{x}_{j}$ be non-decreasing in $\mathbf{x}_{i}$. Recalling now that a composition of convex functions is again convex, then the third condition is ${\partial^2_{\mathbf{x}_0 \mathbf{x}_0} \mathbf{x}_{i}\ge0}$ is satisfied by the recursive nature of the NN.

The two remaining conditions restrict the allowable forms of the NN architecture as shown below. Given the specific format of the NN defined by Eqs.~(\ref{eq_app1_f}), the required derivatives are:
\begin{equation}
	\label{eq_app1_d1st}
	\begin{aligned}
		\partial_{\mathbf{x}_0} \mathbf{x}_{j}&= 
		g'  \left\lbrace  \widetilde{\mathbf{W}}^{x_j}_{x_i y_i}\left[g (\mathbf{W}_{x_i y_i} \mathbf{y}_i+\mathbf{b}_{x_i y_i})\right]\partial_{\mathbf{x}_0} \mathbf{x}_{i}  
		+ \overline{\mathbf{W}}^{x_j}_{x_0 y_i}\left[(\mathbf{W}_{x_i y_i} \mathbf{y}_i+\mathbf{b}_{x_i y_i})\right]  \right\rbrace \\
		\partial^2_{\mathbf{x}_0 \mathbf{x}_0} \mathbf{x}_{j}&= 
		g'' \left\lbrace  \widetilde{\mathbf{W}}^{x_j}_{x_i y_i}\left[g (\mathbf{W}_{x_i y_i} \mathbf{y}_i+\mathbf{b}_{x_i y_i})\right]\partial_{\mathbf{x}_0} \mathbf{x}_{i}  
		+ \overline{\mathbf{W}}^{x_j}_{x_0 y_i}\left[(\mathbf{W}_{x_i y_i} \mathbf{y}_i+\mathbf{b}_{x_i y_i})\right]  \right\rbrace^2 \\   
		&+g'  \left\lbrace  \widetilde{\mathbf{W}}^{x_j}_{x_i y_i}\left[g (\mathbf{W}_{x_i y_i} \mathbf{y}_i+\mathbf{b}_{x_i y_i})\right]\partial^2_{\mathbf{x}_0\mathbf{x}_0} \mathbf{x}_{i} \right\rbrace.	    
	\end{aligned}
\end{equation}
To enforce that the second derivative $\partial^2_{\mathbf{x}_0 \mathbf{x}_0} \mathbf{x}_{j}$ is not negative, the weights $\widetilde{\mathbf{W}}^{x_j}_{x_i y_i}$ must be non-negative. Additionally, the derivatives ${g'\ge0}$ and ${g''\ge0}$ must also be non-negative, which is satisfied if the activation function $g$ is a non-decreasing convex monotonic function.
Under this condition, there are several possible options. However, in the present case, the network must be integrable, which limits the possibilities. In particular, NN describes the strain energy, while the training is performed using stresses, i.e., using the partially differentiated neural network with respect to strain. For this reason, the activation function involving convex variables should have a first derivative that can also be used as an activation function. Among standard activation functions this eventually leaves softplus $g(x)=\log(\exp(x)+1)$ whose first derivative $g'(x)=(1 +\exp(-x ))^{-1} $ is the logistic function as the first choice. Note that the differentiation is not performed with respect to the chirality parameters, so this does not apply to this part of the network. Nevertheless, for simplicity, the softplus activation function was used throughout NN. For other custom-made activation functions, see \cite{linka2023new}.

Note that the connections between the inputs $\mathbf{x}_0, \mathbf{y}_0$ and the hidden layer $\mathbf{x}_1$ are slightly different:
\begin{equation}
	\label{eq_app1_x1x0}
	\begin{array}{l}
		\mathbf{x}_{1}=g \left\lbrace  \mathbf{W}^{x_1}_{x_0 y_0}\left[\mathbf{x}_0 \odot g (\mathbf{W}_{x_0 y_0} \mathbf{y}_0+\mathbf{b}_{x_0 y_0})\right] + \mathbf{W}_{y_0 x_1} \mathbf{y}_0 +\mathbf{b}_{x_1 y_0} \right\rbrace \\
	\end{array}
\end{equation}
see also Fig.~\ref{fig:NN}. The first and second derivatives are:
\begin{equation}
	\label{eq_app1_x1ax0}
	\begin{array}{l}
		\partial_{\mathbf{x}_0}\mathbf{x}_{1}=g' \left\lbrace  \mathbf{W}^{x_1}_{x_0 y_0}\left[ g (\mathbf{W}_{x_0 y_0} \mathbf{y}_0+\mathbf{b}_{x_0 y_0})\right] \right\rbrace \\
		\partial^2_{\mathbf{x}_0\mathbf{x}_0} \mathbf{x}_{1}=g'' \left\lbrace  \mathbf{W}^{x_1}_{x_0 y_0}\left[ g (\mathbf{W}_{x_0 y_0} \mathbf{y}_0+\mathbf{b}_{x_0 y_0})\right] \right\rbrace^2. 
	\end{array}
\end{equation}
Since $g''\ge0$ as explained above, the convexity is preserved in this connection as well.

It must also be demonstrated that the output need not be convex with respect to $\mathbf{y}_0$. Due to the recursive nature of the network, the proof relies on showing that $\partial^2_{\mathbf{y}_{0}\mathbf{y}_{0}} \mathbf{x}_{j}$ can be negative. The first derivative is:
\begin{equation}
	\label{eq_app1_2ndA}
	\begin{aligned}
		\partial_{\mathbf{y}_0} \mathbf{x}_{j} &=g' \left\lbrace \left( \widetilde{\mathbf{W}}^{x_j}_{x_i y_i} \left[ \partial_{\mathbf{y}_0} \mathbf{x}_i \odot g (\mathbf{W}_{x_i y_i} \mathbf{y}_i+\mathbf{b}_{x_i y_i}) + \mathbf{x}_i \odot g' (\mathbf{W}_{x_i y_i} \partial_{\mathbf{y}_0} \mathbf{y}_i)\right]  \right) \right.\\
		\quad \quad \quad \quad & \left. + \overline{\mathbf{W}}^{x_j}_{x_0 y_i} \left[  \mathbf{x}_0 \odot (\mathbf{W}_{x_0 y_i} \partial_{\mathbf{y}_0}\mathbf{y}_i) \right] + \overline{\mathbf{W}}_{y_i} \partial_{\mathbf{y}_0} \mathbf{y}_i  \right\rbrace, 
	\end{aligned}
\end{equation}
where
\begin{equation}
	\label{eq_app1_y0_2ndA}
	\begin{aligned}
		\partial_{\mathbf{y}_{0}} \mathbf{y}_{j}&=g'\overline{\mathbf{W}}_{y_i y_j}\partial_{\mathbf{y}_{0}} \mathbf{y}_i \\
		\partial^2_{\mathbf{y}_{0}\mathbf{y}_{0}} \mathbf{y}_{j}&=g''\overline{\mathbf{W}}_{y_i y_j}\partial_{\mathbf{y}_{0}} \mathbf{y}_i+g'\overline{\mathbf{W}}_{y_i y_j}\partial^2_{\mathbf{y}_{0}\mathbf{y}_{0}} \mathbf{y}_i .
	\end{aligned}
\end{equation}

The second derivative leads to a lengthy form in which several terms must be summed. It can be easily obtained and is not shown here. To show that the second derivative can be negative, it is sufficient to show that at least one term is nonconvex. Consider only the last term of the first derivative Eq.~(\ref{eq_app1_2ndA}). Differentiation yields the second derivative as $g' \overline{\mathbf{W}}_{y_i} \partial^2_{\mathbf{y}_{0}\mathbf{y}_{0}} \mathbf{y}_{i}$, where no positivity constraint is applied to the weights, which together with Eq.~(\ref{eq_app1_y0_2ndA}) provides a term that is not convex, which consequently shows that the NN output is not convex with respect to $\mathbf{y}_0$.

\section{Positivity Constraint and Activation Functions}
\label{sec:app2}
As discussed in \ref{sec:app1}, a non-negativity constraint should be introduced for some of the weights. The present choice is the same as in \cite{amos2017input,huang2022variational}. For each element $W_{ij}$ of a weight matrix $\mathbf{W}$, the following transformation $\mathbf{W}\rightarrow\widetilde{\mathbf{W}}$ is used:
\begin{equation}
	\label{eq_app2_Wij}
	\begin{aligned}
		\widetilde{W}_{ij}=W_{ij} + \exp{(-\gamma)},  \quad \forall W_{ij} \in \mathbf{W} \ge 0 \\
		\widetilde{W}_{ij} = \exp{(W_{ij}-\gamma)},  \quad \forall {W}_{ij} \in \mathbf{W}  <0, \\
	\end{aligned}
\end{equation}
where $\gamma=5$. For other possibilities, see \cite{thakolkaran2022nn, as2022mechanics}.

\section{Neural Network for Approximation of SWCNT Diameters}
\label{sec:app3}
As discussed in Sec.~\ref{sec_abaqus}, the standard Abaqus UMAT implementation works fine unless the analysis involves larger deformations. In such a case, one should account for the correct value of the current SWCNT diameter rather than the value resulting from the incompressibility constraint used by Abaqus. To solve this issue, another NN was developed that approximates the SWCNT diameter. This NN is also chosen to be of the PICINN type (although integrability is not required) and has a similar architecture to the NN used for strain energy and described in Sec.~\ref{sec_ML_model}. However, the strain energy is now replaced by the diameter and the training was not performed on derivatives, but on actual values of the diameter as obtained from MD. This network is larger, while the dimensionality of the inputs/outputs is now defined as $n_{x0}=n_{x3}=1$, $n_{y0}=2$, $n_{y}=25$, $n_{x1}=n_{x2}=20$. The learning rate was increased to $0.001$. All other hyperparameters remained the same.

\begin{figure}
	\centering
	\includegraphics[scale=1.0]{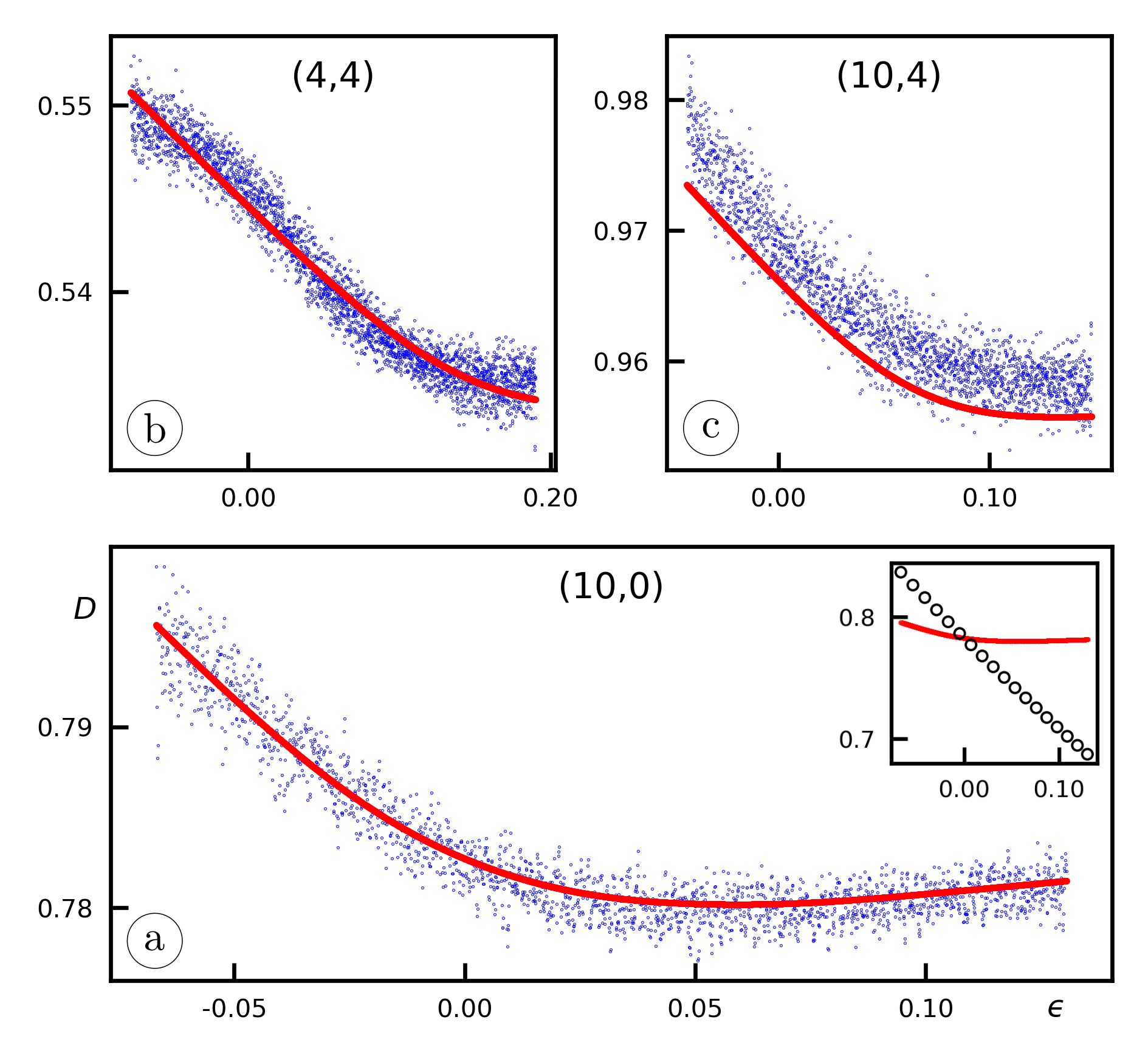}
	\caption{Comparison of diameter vs. true strain curves as obtained from MD (blue) and their NN approximations (red). (a) A typical approximation, (10, 0) SWCNT; the inset shows the comparison of the actual diameter (red) with the diameter as calculated assuming incompressibility (black dots); (b) an example of the deviation from convexity at the beginning and at the end of the $D-\epsilon$ curve, (4, 4) SWCNT; (c) approximation error, (10, 4) SWCNT.} 
	\label{fig:diameters}
\end{figure}

In this way, the diameter will be a convex function of the strain. This approach has the advantage of obtaining smooth curves, without negative gradients, which are present in the case of diameter calculation \citep{canadija2021deep}. A selection of typical ones as well as some results with small discrepancies is provided in Fig.~\ref{fig:diameters}. It can be argued that restricting the diameter to be convex in strain is not physical, but visual inspection of the results from MD has shown that this leads to a reasonable approximation of the diameter (Fig.~\ref{fig:diameters}a), with only a very small number of exceptions in parts of some curves. These notably include very small diameters (Fig.~\ref{fig:diameters}b), and since the approximation of such a cross section as a circular ring is already questionable, the additional error introduced in this way is very small, in any case much smaller than the assumption of the incompressibility constraint (inset in Fig.~\ref{fig:diameters}a). Only six such curves out of 818 were found, most of them for armchair configurations up to (7, 7). In some rarer cases, the approximate curves are also offset, Fig.\ref{fig:diameters}c, although they capture the correct shape of the curve. Using visual inspection 38 such curves were identified. Although this may appear to be a larger error, a simple calculation shows that this error is on average less than 0.5\%.

\bigskip

\noindent
\printcredits

\bibliographystyle{cas-model2-names}

\bibliography{SmallSizeParameter}

\end{document}